\documentclass[11pt]{article}
\usepackage[a4paper,margin=2.3cm]{geometry}
\usepackage{amsmath,amssymb,amsthm}
\usepackage{mathtools}
\usepackage{physics}
\usepackage{bm}
\usepackage{graphicx}
\usepackage{enumitem}
\usepackage{booktabs}
\usepackage{titlesec}
\usepackage{xcolor}
\usepackage{tikz}
\usepackage{authblk}
\usepackage[font=small,labelfont=bf]{caption}
\usepackage[colorlinks=true,linkcolor=red,urlcolor=red,citecolor=red]{hyperref}

\makeatletter
\renewcommand*{\@makefnmark}{\textcolor{red}{\@thefnmark}}
\makeatother

\titleformat{\section}{\normalfont\Large\bfseries}{\thesection}{1em}{}
\titleformat{\subsection}{\normalfont\large\bfseries}{\thesubsection}{1em}{}

\graphicspath{{figures/}}

\title{\textbf{Theory of Magic Angles in Twisted Bilayer Graphene: from Non-Abelian Gauge Fields to Flat Bands}}

\author[1,2,3]{Leonardo A. Navarro-Labastida\thanks{\href{mailto:ln2622@nyu.edu}{ln2622@nyu.edu}}}
\author[1]{Gerardo G. Naumis\thanks{\href{mailto:naumis@fisica.unam.mx}{naumis@fisica.unam.mx}}}
\affil[1]{Instituto de F\'{i}sica, Universidad Nacional Aut\'{o}noma de M\'{e}xico (UNAM), 01000 M\'{e}xico D.F., M\'{e}xico}
\affil[2]{Arts and Sciences, NYU Shanghai, Shanghai 200124, China}
\affil[3]{NYU-ECNU Institute of Physics at NYU Shanghai, Shanghai 200062, China}

\date{\today}

\newcommand{\ii}{\mathrm{i}}
\renewcommand{\vec}[1]{\bm{#1}}

\begin{document}

\maketitle

\begin{abstract}
\noindent
At a discrete set of ``magic'' twist angles, twisted bilayer graphene develops two bands at charge neutrality that are almost perfectly flat. We present a unified theory of this phenomenon within the chiral continuum model. The central object is the square of the chiral Hamiltonian, a $2\times2$ Schr\"odinger operator for an electron moving in a non-Abelian $SU(2)$ pseudo-magnetic field generated by the interlayer tunneling. From it we derive an energy sum rule in which the interlayer coupling splits into a bounded overlap channel and an unbounded current channel; their different scaling with the coupling $\alpha$ explains why the first magic angle is qualitatively different from the rest. The asymptotic rule $\alpha_{m+1}-\alpha_m\to3/2$ follows from a rescaling of the moir\'e potential together with its three-cell magnetic periodicity, with the two parity families of zero modes each repeating with period $3$ and interleaved with each other. Near the AA point the zero-mode problem becomes a lowest-Landau-level problem in an effective field $B_{\rm eff}=3\alpha$, so that the high-order zero modes are coherent Landau states of width $1/\sqrt{3\alpha}$ whose guiding centres converge to $\pm(\pi/3)\vec q_\mu$, with kinetic and confinement energies in equipartition. Deforming the coupling into an Abelian one removes the magic-angle sequence, which shows that the non-Abelian structure is essential. Between consecutive magic angles, band inversions at $\Gamma$ and $M$ produce phases with Chern numbers $C=\pm2$, and the quantum metric reaches a sharp maximum at $\theta\approx0.43^\circ$, identifying a promising regime for correlated and topological phases beyond the first magic angle.
\end{abstract}

\tableofcontents

\section{Introduction}
\label{sec:intro}

\subsection{Why twist two graphene layers?}

If two graphene sheets are stacked with a small relative rotation $\theta$, the two triangular lattices drift in and out of registry and form a moir\'e pattern with period $L_m\approx a_0/\theta$, where $a_0$ is the graphene lattice constant (Fig.~\ref{fig:geometry}a). At $\theta\approx1^\circ$ the moir\'e period is about fifty times $a_0$. The moir\'e Brillouin zone shrinks accordingly, and the kinetic energy scale of the low-energy electrons drops with it \cite{andrei2020,balents2020}. At certain twist angles the two bands closest to charge neutrality become almost dispersionless. Their bandwidth then falls well below the Coulomb scale $U_C\sim e^2/(4\pi\epsilon L_m)$, and interactions dominate. This is the origin of the correlated insulators, superconductivity and Chern insulators seen in twisted bilayer graphene (TBG) \cite{cao2018a,cao2018b,park2021}.

These notes deal only with the single-particle problem. We want to understand why the magic angles form the sequence they do, what the flat-band wavefunctions look like, and what happens in between two magic angles. Most of the answers come from one object, the square of the chiral continuum Hamiltonian, and the notes are organized around it. The material is based on our papers \cite{naumis2021,navarro2022,navarro2023a,navarro2024,navarro2026a,navarro2023b,navarrosusy2025}; here we try to present it in a more pedagogical and self-contained way.

\subsection{The continuum Hamiltonian}

\begin{figure}[t]
\centering
\includegraphics[width=\linewidth]{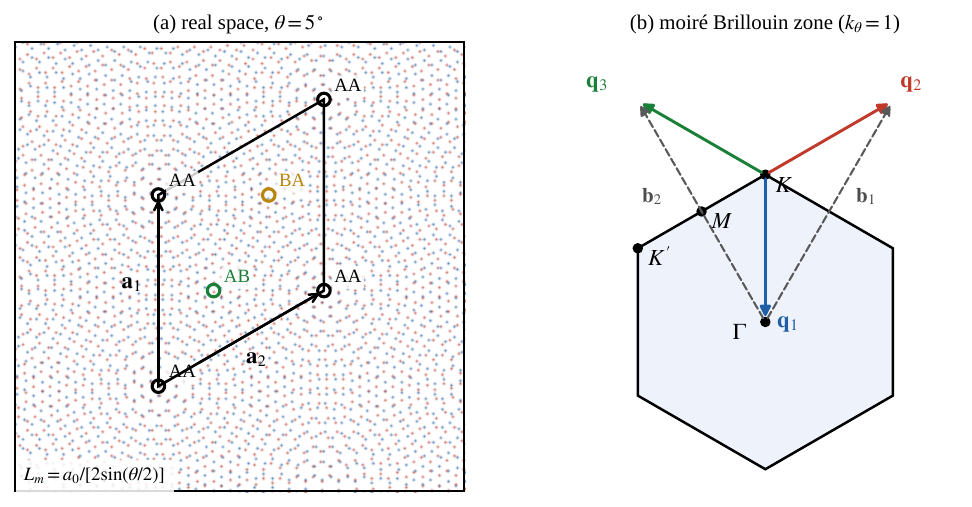}
\caption{Geometry of TBG. (a) Two honeycomb lattices (blue and red) rotated by $\theta=5^\circ$. The moir\'e cell spanned by $\vec a_1,\vec a_2$ has AA regions at its corners, where the atoms of both layers sit on top of each other, and AB and BA regions inside, where one sublattice of one layer sits on top of the other sublattice of the other layer. (b) Moir\'e Brillouin zone in units of $k_\theta$. The vectors $\vec q_\nu$ join the Dirac points of the two layers; $\vec b_{1,2}$ are reciprocal lattice vectors.}
\label{fig:geometry}
\end{figure}

At long wavelengths TBG is described by two rotated Dirac cones coupled through a smooth, moir\'e-periodic tunneling \cite{bistritzer2011}. We use the four-component spinor $\Phi(\vec r)=(\psi_1,\psi_2,\chi_1,\chi_2)^T$, where $1,2$ labels the layer and $\psi$, $\chi$ the two graphene sublattices. In the sublattice-block basis $(\Psi,\chi)$, with $\Psi=(\psi_1,\psi_2)^T$ and $\chi=(\chi_1,\chi_2)^T$, the Bistritzer--MacDonald (BM) Hamiltonian reads
\begin{equation}
H = \begin{pmatrix} M & D^\dagger \\ D & M \end{pmatrix},
\qquad
D(\vec r) = \begin{pmatrix} -\ii\bar\partial & \alpha U_1(\vec r) \\ \alpha U_1(-\vec r) & -\ii\bar\partial \end{pmatrix},
\label{eq:BMHam}
\end{equation}
with $\bar\partial=\partial_x+\ii\partial_y$ and $\partial=\partial_x-\ii\partial_y$. The tunneling functions are $U_m(\vec r)=\sum_{\nu=1}^3 e^{\ii m\phi(\nu-1)}e^{-\ii \vec q_\nu\cdot\vec r}$, with $\phi=2\pi/3$ and $m=0,\pm1$. The three vectors
\begin{equation}
\vec q_1 = k_\theta(0,-1),\quad
\vec q_2 = k_\theta\Big(\tfrac{\sqrt3}{2},\tfrac12\Big),\quad
\vec q_3 = k_\theta\Big(-\tfrac{\sqrt3}{2},\tfrac12\Big)
\label{eq:qvectors}
\end{equation}
connect the Dirac points of the two layers (Fig.~\ref{fig:geometry}b). Here $k_\theta=2k_D\sin(\theta/2)$ and $k_D=4\pi/3a_0$. The model has a single dimensionless parameter,
\begin{equation}
\alpha = \frac{w_{1}}{v_0 k_\theta},
\label{eq:alphadef}
\end{equation}
where $w_1\equiv w_{AB}\approx110$~meV is the interlayer hopping in the AB/BA regions and $v_0$ is the Fermi velocity of monolayer graphene, $v_0=19.81\,\mathrm{eV}/(2k_D)$. For small angles this gives $\theta\approx0.636^\circ/\alpha$. From now on we set $v_0=k_\theta=1$, so that a smaller twist angle simply means a larger $\alpha$. The moir\'e reciprocal vectors are $\vec b_{1,2}=\vec q_{2,3}-\vec q_1$ and $\vec b_3=\vec q_3-\vec q_2$, the moir\'e lattice vectors are $\vec a_{1,2}=(4\pi/3)(\pm\sqrt3/2,1/2)$, and the high-symmetry points we use are listed in Table~\ref{tab:hspoints}.

\begin{table}[h]
\centering
\caption{High-symmetry points of the moir\'e Brillouin zone in units of $k_\theta$ (see Fig.~\ref{fig:geometry}b).}
\label{tab:hspoints}
\begin{tabular}{@{}lcccc@{}}
\toprule
Point & $\Gamma$ & $K$ & $K'$ & $M$ \\ \midrule
Position & $\vec q_1$ & $(0,0)$ & $-\vec q_1$ & $-\vec q_1/2$\\
\bottomrule
\end{tabular}
\end{table}

The block
\begin{equation}
M(\vec r) = \kappa\begin{pmatrix}0&\alpha U_0(\vec r)\\\alpha U_0(-\vec r)&0\end{pmatrix}, \qquad \kappa=\frac{w_{AA}}{w_{AB}},
\label{eq:Mkappaintro}
\end{equation}
contains the same-sublattice (AA) tunneling. It is diagonal in the sublattice blocks and off-diagonal in layer space, so it couples $A_1\leftrightarrow A_2$ and $B_1\leftrightarrow B_2$.

The Hamiltonian is invariant under $C_{3z}$, $C_{2x}$ and $C_{2z}\mathcal T$, where $\mathcal T$ is time reversal. In the conventions used here,
\begin{equation}
C_3 = e^{\ii 2\pi\sigma_z/3}\ (\vec r\to R_{2\pi/3}\vec r), \qquad
C_2\mathcal T = \sigma_x\mathcal K\ (\vec r\to-\vec r), \qquad
\mathcal M = \eta_x\sigma_x\ (y\to-y),
\label{eq:symmetries}
\end{equation}
with $\sigma$ and $\eta$ Pauli matrices in sublattice and layer space and $\mathcal K$ complex conjugation. $C_3$ is responsible for the threefold pattern of every real-space plot below. $C_{2z}\mathcal T$ acts within one valley and protects the Dirac points at $K$ and $K'$, while $\mathcal T$ maps one valley onto the other.

\subsection{The chiral limit and the magic angles}
\label{sec:chirallimit}

Lattice relaxation shrinks the AA regions and enlarges the AB/BA ones, so that $w_{AA}<w_{AB}$ in real samples ($\kappa\approx0.7$--$0.8$) \cite{nam2017,koshino2018}. Tarnopolsky, Kruchkov and Vishwanath (TKV) observed that the limit $w_{AA}=0$ is very special \cite{tarnopolsky2019}. In this chiral limit $H$ anticommutes with $\Gamma=\mathrm{diag}(1,1,-1,-1)$ and becomes
\begin{equation}
H = \begin{pmatrix} 0 & D^\dagger(-\vec r) \\ D(\vec r) & 0 \end{pmatrix},
\qquad
D(\vec r) = \begin{pmatrix} -\ii\bar\partial & \alpha U(\vec r) \\ \alpha U(-\vec r) & -\ii\bar\partial \end{pmatrix},
\label{eq:chiralH}
\end{equation}
with $U(\vec r)=e^{-\ii\vec q_1\cdot\vec r}+e^{\ii\phi}e^{-\ii\vec q_2\cdot\vec r}+e^{-\ii\phi}e^{-\ii\vec q_3\cdot\vec r}$. Together with time reversal and particle-hole symmetry,
\begin{equation}
TH(\vec k)T^{-1}=H(-\vec k), \qquad CH(\vec k)C^{-1}=-H(-\vec k), \qquad \Gamma H(\vec k)\Gamma^{-1}=-H(\vec k),
\label{eq:susy-symm}
\end{equation}
the chiral symmetry allows us to write $H(\vec k)=\big(\begin{smallmatrix}0&Q^\dagger(\vec k)\\Q(\vec k)&0\end{smallmatrix}\big)$, with $Q(\vec k)$ the Bloch form of $D(\vec r)$. This is the structure of supersymmetric quantum mechanics, with $Q$ playing the role of the supercharge. We come back to this in Sec.~\ref{sec:squaring}.

\begin{figure}[t]
\centering
\includegraphics[width=\linewidth]{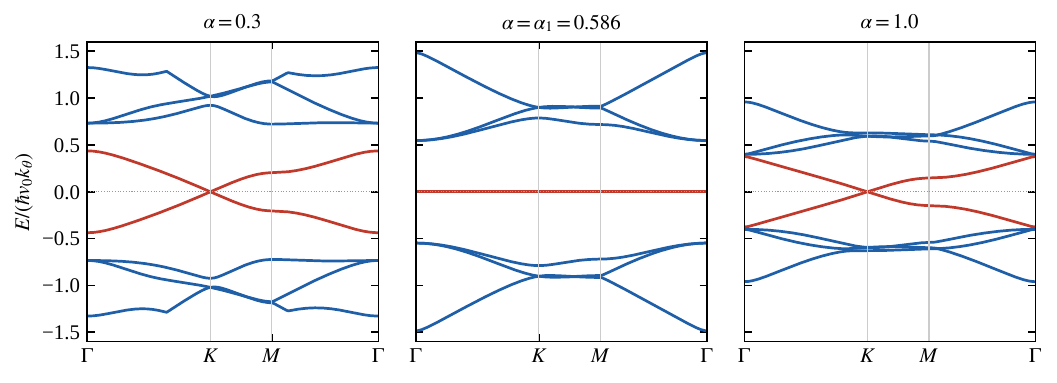}
\caption{Bands of the chiral Hamiltonian~\eqref{eq:chiralH} along $\Gamma$--$K$--$M$--$\Gamma$, below, at, and above the first magic angle. The two central bands are shown in red. At $\alpha_1$ they are exactly flat.}
\label{fig:bands}
\end{figure}

In the chiral model there is a discrete set of couplings $\alpha_m$ at which the two central bands are exactly flat and pinned at $E=0$ \cite{tarnopolsky2019,khalaf2019}. Figure~\ref{fig:bands} shows what happens as $\alpha$ crosses the first one. Below $\alpha_1$ the central bands form two Dirac cones at $K$ and $K'$; at $\alpha_1$ they collapse onto $E=0$ over the whole Brillouin zone; above $\alpha_1$ they disperse again, with a Fermi velocity of opposite sign. Table~\ref{tab:magicangles} lists the first nine magic angles. For large $m$ the spacing between consecutive values approaches $3/2$,
\begin{equation}
\alpha_{m+1}-\alpha_m \;\longrightarrow\; \frac{3}{2}, \qquad m\gg1.
\label{eq:32rule}
\end{equation}
The explanation of this rule is the subject of Sec.~\ref{sec:rescaling}. The table also lists the parity $\mu_\alpha=\pm1$ of the zero mode at $\Gamma$, defined in Sec.~\ref{sec:squaringinversion}; it alternates from one magic angle to the next, and this turns out to be the key to the factor $1/2$ in Eq.~\eqref{eq:32rule}.

\begin{table}[h]
\centering
\caption{First nine magic angles of the chiral model, their spacing, and the parity $\mu_\alpha$ of the $\Gamma$-point zero mode ($\psi_2(\vec r)=\ii\mu_\alpha\psi_1(-\vec r)$). Values computed with the method of Appendix~\ref{app:bloch}; they agree with Ref.~\cite{tarnopolsky2019}. The spacing between magic angles of the same parity is $\alpha_{m+2}-\alpha_m=3.17,\,3.06,\,3.04,\dots,3.03$.}
\label{tab:magicangles}
\begin{tabular}{@{}cccccccccc@{}}
\toprule
$m$ & 1 & 2 & 3 & 4 & 5 & 6 & 7 & 8 & 9\\ \midrule
$\alpha_m$ & 0.586 & 2.221 & 3.751 & 5.276 & 6.795 & 8.313 & 9.829 & 11.345 & 12.861 \\
$\alpha_{m+1}-\alpha_m$ & 1.636 & 1.530 & 1.525 & 1.518 & 1.518 & 1.516 & 1.516 & 1.515 & 1.515 \\
$\mu_\alpha$ & $+$ & $-$ & $+$ & $-$ & $+$ & $-$ & $+$ & $-$ & $+$ \\
\bottomrule
\end{tabular}
\end{table}

\subsection{Energy scales}

A first estimate of $\alpha_1$ comes from perturbation theory around the Dirac point of the BM model \cite{bistritzer2011}. For $w_{AA}=w_{AB}$ it gives the renormalized Fermi velocity
\begin{equation}
\frac{v^*}{v_0} = \frac{1-3\alpha^2}{1+6\alpha^2},
\label{eq:vstar}
\end{equation}
which vanishes at $\alpha=1/\sqrt3\approx0.577$, close to the exact $\alpha_1=0.586$. The mechanism is already visible here: interlayer tunneling cancels the intralayer kinetic energy. The estimate keeps only the first shell of plane waves, so it cannot say anything about the higher magic angles.

The reason all this matters experimentally is a competition of scales. The moir\'e period is
\begin{equation}
L_m = \frac{a_0}{2\sin(\theta/2)} \approx \frac{a_0}{\theta}.
\label{eq:moireperiod}
\end{equation}
The bandwidth of the central bands scales as $W\sim\hbar v^*/L_m$, while the Coulomb energy $U_C\sim e^2/(4\pi\epsilon_0\epsilon_r L_m)\sim10$--$20$~meV. Both decrease as $L_m$ grows, but $W$ vanishes at a magic angle while $U_C$ does not, so the ratio $U_C/W$ diverges there. This is why the correlated phases appear in a narrow window around $\theta\approx1.1^\circ$ \cite{yankowitz2019,lu2019,stepanov2020}. The rest of these notes deal with the statement ``$W=0$ at $\alpha_m$'' and with what surrounds it.

\section{Zero Modes and the Squared Hamiltonian}
\label{sec:squaring}

\subsection{Flat-band wavefunctions}

Inside the flat band, the Bloch states factorize as \cite{tarnopolsky2019}
\begin{equation}
\begin{pmatrix}\psi_{\vec k,1}(\vec r)\\ \psi_{\vec k,2}(\vec r)\end{pmatrix}
= f_{\vec k}(z)\begin{pmatrix}\psi_{K,1}(\vec r)\\ \psi_{K,2}(\vec r)\end{pmatrix},
\qquad z=x+\ii y,
\label{eq:flatbandfactor}
\end{equation}
where $f_{\vec k}(z)$ is a ratio of Jacobi theta functions fixed by the Bloch boundary conditions. Its poles are cancelled by the zero of $\psi_K(\vec r)$, which is what makes the construction possible only at magic angles. All the $\vec k$ dependence of the flat band sits in this analytic prefactor. Near the first magic angle, perturbation theory gives the layer densities at $K$ \cite{navarro2022}
\begin{align}
\rho_{K,1}(\vec r) &= 1+\frac{4\alpha^2}{\sqrt3}\sum_{\mu=1}^{3}\sin\!\big(\phi+(-1)^{\mu-1}\vec b_\mu\cdot\vec r\big) + \mathcal O(\alpha^4),
\label{eq:rhoK1}
\\
\rho_{K,2}(\vec r) &= \alpha^2\Big(3+2\sum_{\mu=1}^3\cos(\vec b_\mu\cdot\vec r)\Big).
\label{eq:rhoK2}
\end{align}
The second density has the same form as the confinement potential of Eq.~\eqref{eq:confpot} below.  This is an early hint that the first magic angle behaves differently.

\subsection{Squaring the chiral Hamiltonian}

The most useful single step in this theory is to square $H$. For $H=\big(\begin{smallmatrix}0&Q^\dagger\\Q&0\end{smallmatrix}\big)$ we get $H^2=\mathrm{diag}(Q^\dagger Q,\,QQ^\dagger)$. The two blocks are supersymmetric partners: they share every nonzero eigenvalue and differ only in their zero modes. Yoshida, Mizoguchi, Kuno and Hatsugai showed that for Hamiltonians of this symmetry class both blocks carry the same $\mathbb Z_2$ invariant \cite{yoshida2021}. For TBG, squaring turns the $4\times4$ Dirac problem into a $2\times2$ Schr\"odinger-like operator \cite{naumis2021},
\begin{equation}
H^2 = \begin{pmatrix} -\nabla^2+\alpha^2|U(-\vec r)|^2 & \alpha A^\dagger(\vec r) \\ \alpha A(\vec r) & -\nabla^2+\alpha^2|U(\vec r)|^2 \end{pmatrix},
\label{eq:H2reduced}
\end{equation}
with the confinement potential
\begin{equation}
|U(\vec r)|^2 = 3 + 2\cos(\vec b_1\cdot\vec r-\phi) + 2\cos(\vec b_2\cdot\vec r+\phi) + 2\cos(\vec b_3\cdot\vec r+2\phi)
\label{eq:confpot}
\end{equation}
and the interlayer operator
\begin{equation}
A^\dagger(\vec r) = -\ii\sum_{\mu=1}^{3} e^{-\ii \vec q_\mu\cdot\vec r}\big(2\hat q_\mu^\perp\cdot\nabla+1\big),
\qquad \hat q_1^\perp=(1,0),\ \hat q_2^\perp=\Big(-\tfrac12,\tfrac{\sqrt3}{2}\Big),\ \hat q_3^\perp=\Big(-\tfrac12,-\tfrac{\sqrt3}{2}\Big).
\label{eq:Aoperator}
\end{equation}
Each term has a clear meaning: a kinetic energy $-\nabla^2$, a triangular confinement potential for each layer (Fig.~\ref{fig:density}a,b), and an interlayer coupling $A(\vec r)$ that contains a gradient. The potential $|U(\vec r)|^2$ vanishes at AA and AB and reaches its maximum, $9$, at BA; for the other layer the roles of AB and BA are exchanged.

Squaring has a long history in bipartite lattices. It maps a honeycomb lattice onto an effective triangular lattice with renormalized hoppings \cite{naumis2007,barrios2011}, and it folds the spectrum around $E=0$ so that the states at $\pm E$ become degenerate states at $E^2$. After squaring, each sublattice sector is a separate problem, and solving one of them gives the other by symmetry (Sec.~\ref{sec:squaringinversion}). A zero mode of $H$ is then a ground state of $H^2$. On a triangular lattice such a state is frustrated: it cannot alternate its phase between all neighbouring sites. This is the same frustration mechanism that produces highly degenerate confined states in other bipartite problems, from doped graphene to the Penrose tiling \cite{naumis1994}, and we find it a useful way to think about why flat bands appear at all.

Writing the off-diagonal block with the layer Pauli matrices $\tau_j$,
\begin{equation}
\begin{gathered}
\hat A_x = A_{1,x}\tau_1+A_{2,x}\tau_2,\qquad \hat A_y = A_{1,y}\tau_1+A_{2,y}\tau_2, \\
A_{1,\nu}=\sum_\mu \cos(\vec q_\mu\cdot\vec r)\,\hat q_\mu^{\perp,\nu}, \qquad
A_{2,\nu}=\sum_\mu \sin(\vec q_\mu\cdot\vec r)\,\hat q_\mu^{\perp,\nu},
\end{gathered}
\label{eq:SU2A}
\end{equation}
we find that the two components do not commute, not even at the same point. With $\tau_j\tau_k=\delta_{jk}\tau_0+\ii\epsilon_{jkl}\tau_l$,
\begin{equation}
[\hat A_x(\vec r),\hat A_y(\vec r)] = 2\ii\big(A_{1,x}A_{2,y}-A_{2,x}A_{1,y}\big)\tau_3,
\label{eq:commutatorexplicit}
\end{equation}
which is nonzero except on special lines of the moir\'e cell. So $\hat{\vec A}$ is a non-Abelian $SU(2)$ gauge field \cite{sanjose2012}. With the canonical momentum $\hat{\vec\Pi}=\vec p+\alpha\hat{\vec A}$, the squared Hamiltonian takes the form \cite{naumis2021,navarro2022,navarrosusy2025}
\begin{equation}
H^2 = \hat{\vec\Pi}\cdot\hat{\vec\Pi} - \ii(\hat{\vec\Pi}\times\hat{\vec\Pi})
= -\nabla^2 + \alpha^2\hat A^2 - 2\ii\alpha(\hat A\cdot\nabla) + \alpha \hat F_{xy},
\qquad
\hat F_{xy} = \partial_x\hat A_y - \partial_y \hat A_x - \ii\alpha[\hat A_x,\hat A_y].
\label{eq:H2enlightening}
\end{equation}
This is an electron in a non-Abelian pseudo-magnetic field. The term $\alpha^2\hat A^2$ is a magnetic confinement energy, $-2\ii\alpha(\hat A\cdot\nabla)$ is the orbital coupling between field and momentum, and $\alpha\hat F_{xy}$ is a Zeeman-like coupling to the layer pseudospin. The commutator inside $\hat F_{xy}$ has no counterpart in an ordinary quantum Hall problem. Since the field is built from the interlayer tunneling, time-reversal symmetry is never broken, even though the pseudo-fields are complex. Most of what follows comes from looking at Eq.~\eqref{eq:H2enlightening} in different limits.

\subsection{Confinement, commutator and inversion symmetry}
\label{sec:squaringinversion}

It is convenient to split the norm and the commutator of the complex vector potential $\vec A_\pm(\vec r)=\sum_\mu e^{\pm\ii\vec q_\mu\cdot\vec r}\hat q_\mu^\perp$ \cite{navarro2024},
\begin{equation}
A^2(\vec r) = 3-\sum_{\nu=1}^3 \cos(\vec b_\nu\cdot\vec r),
\qquad
\Delta(\vec r) = \ii[A_x,A_y] = \sqrt3 \sum_{\nu=1}^3 (-1)^\nu \sin(\vec b_\nu\cdot\vec r),
\label{eq:A2Delta}
\end{equation}
so that the diagonal blocks of $H^2$ read $-\nabla^2+\alpha^2(A^2\mp\Delta)$. Here $A^2$ is even in $\vec r$ and $\Delta$ is odd. Much of Sec.~\ref{sec:localization} reduces to asking which of the two dominates in a given region of the cell.

The chiral model also has an intravalley inversion symmetry, $\hat{\mathcal I}=\sigma_z\tau_y$ with $\hat{\mathcal I}H(\vec r)\hat{\mathcal I}^\dagger=H(-\vec r)$ \cite{wang2021a}. It survives squaring, $\hat{\mathcal I}H^2(\vec r)\hat{\mathcal I}^\dagger=H^2(-\vec r)$. At the $\Gamma$ point it fixes the relation between the two layers of a zero mode,
\begin{equation}
\psi_2(\vec r)=\ii\mu_\alpha\psi_1(-\vec r), \qquad \mu_\alpha=\pm1,
\label{eq:parityrelation}
\end{equation}
where $\mu_\alpha$ is the parity of the magic angle \cite{tarnopolsky2019}. Parity (together with chirality) also acts as a selection rule in related moir\'e systems; for instance, in hybrid moir\'e--one-dimensional superlattices it determines which Dirac masses can be opened \cite{tan2026}. Once $\psi_1$ is known, $\psi_2$ follows. In practice this reduces every two-component zero-mode problem at $\Gamma$ to a one-component one, which is why the expansions of Appendix~\ref{app:pert} only track $\psi_1$.

\begin{figure}[t]
\centering
\includegraphics[width=\linewidth]{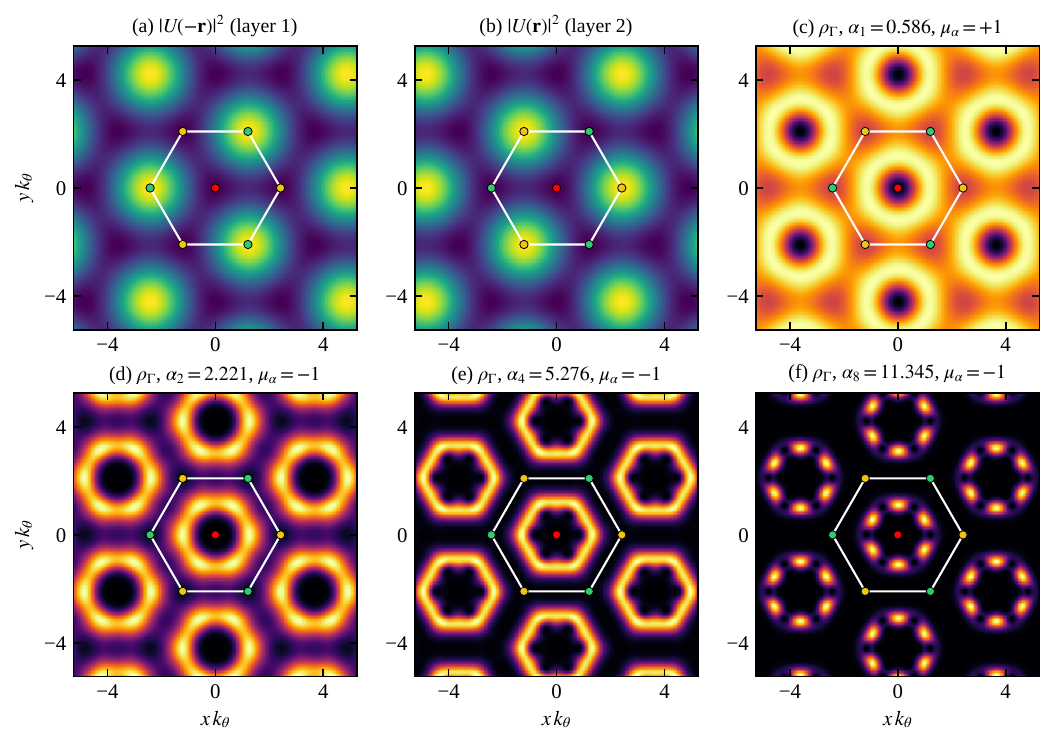}
\caption{(a),(b) Confinement potentials of the two layers in Eq.~\eqref{eq:H2reduced}. (c)--(f) Electronic density $\rho_\Gamma=|\psi_1|^2+|\psi_2|^2$ of the $\Gamma$-point zero mode at the magic angles $\alpha_1$, $\alpha_2$, $\alpha_4$ and $\alpha_8$, normalized to its maximum. The white hexagon is the moir\'e Wigner--Seitz cell centred at AA (red dot); yellow and green dots mark BA and AB. As $m$ grows the density is pushed away from AA and concentrates on a ring of radius close to $\pi/3\approx1.05$, breaking into six spots (Sec.~\ref{sec:localization}).}
\label{fig:density}
\end{figure}

Figure~\ref{fig:density} shows what the zero modes look like. At $\alpha_1$ the density at $\Gamma$ is spread over the cell, with a minimum at AA surrounded by a broad ring. At higher magic angles the ring gets thinner and moves towards the points $\pm(\pi/3)\vec q_\nu$, and at $\alpha_8$ it has broken into six localized spots. The rest of the notes try to explain this evolution.

\section{Energy Balance, Interlayer Currents and the 3/2 Rule}
\label{sec:qhe}

In this section we look at the magic angles from the point of view of energy. Taking expectation values of $H^2$ separates the kinetic, confinement and interlayer contributions, and the way they balance tells us a lot about the zero modes. The second part of the section explains the $3/2$ rule. The analogy with the quantum Hall effect used here was suggested in Ref.~\cite{navarro2023b}. We stress that the pseudo-magnetic field is internal to $H^2$ and there is no external field. It should not be confused with the Landau levels and Hofstadter spectra of TBG in a real magnetic field \cite{moon2012,moon2013,crosse2020}.

\subsection{An energy sum rule}

Integrating $\Psi_{\vec k}^\dagger H^2\Psi_{\vec k}=E^2(\vec k)\,\Psi_{\vec k}^\dagger\Psi_{\vec k}$ over the moir\'e cell gives \cite{navarro2022}
\begin{equation}
\sum_{j=1}^{2}\big(\expval{T_j}+\expval{V_j}\big) + \expval{\mathcal A} = E^2(\vec k),
\label{eq:sumrule}
\end{equation}
with the kinetic and confinement energies of each layer
\begin{equation}
\expval{T_j} = -\int_m \psi_{\vec k,j}^*(\vec r)\,\nabla^2\psi_{\vec k,j}(\vec r)\,\dd^2r,
\qquad
\expval{V_j} = \alpha^2\int_m |U(\mp\vec r)|^2\,\rho_{\vec k,j}(\vec r)\,\dd^2r,
\label{eq:TVdef}
\end{equation}
where the upper (lower) sign is for layer 1 (2). Since $0\le|U|^2\le9$ and its cell average is $3$, a normalized state has $0\le\expval{V}\le9\alpha^2$, and a uniform density would give $\expval{V}=3\alpha^2$. The interlayer term is $\expval{\mathcal A}=\expval{A^\dagger}_{12}+\expval{A}_{21}$, with
\begin{equation}
\expval{A^\dagger}_{12} = \alpha\int_m \psi_{\vec k,1}^*\,A^\dagger(\vec r)\,\psi_{\vec k,2}\,\dd^2r,
\qquad
\expval{A}_{21} = \alpha\int_m \psi_{\vec k,2}^*\,A(\vec r)\,\psi_{\vec k,1}\,\dd^2r.
\label{eq:Adaggerdef}
\end{equation}
Each of the two terms can be complex; their sum is real. The interlayer operator has two pieces with different physics. The gradient piece,
\begin{equation}
A_g(\vec r) = -2\ii\alpha \sum_\mu \hat q_\mu^\perp \cdot \big(e^{\ii\vec q_\mu\cdot\vec r}\psi_1\nabla\psi_2^* + e^{-\ii\vec q_\mu\cdot\vec r}\psi_2^*\nabla\psi_1\big),
\label{eq:Agdef}
\end{equation}
is an interlayer current, $\expval{A_g}=2\alpha\sum_\mu\hat q_\mu^\perp\cdot\expval{\tilde{\vec j}_{12}(\vec q_\mu)+\tilde{\vec j}_{21}(-\vec q_\mu)}$ with $\vec j_{12}=\tfrac{\ii e\hbar}{2m}(\psi_1\nabla\psi_2^*-\psi_2^*\nabla\psi_1)$. On the lattice it corresponds to the second-neighbour current $\ii(a_{2,j}^\dagger a_{1,i}-a_{2,i}^\dagger a_{1,j})$ between layers on the same sublattice. The other piece,
\begin{equation}
A_f(\vec r) = \ii\alpha \sum_\mu \big(\psi_2^*\psi_1 e^{\ii\vec q_\mu\cdot\vec r} - \psi_1^*\psi_2 e^{-\ii\vec q_\mu\cdot\vec r}\big),
\label{eq:Afdef}
\end{equation}
contains no derivative. It measures the overlap between the layers and is bounded, $|\expval{A_f}|\le6\alpha$. The difference between these two channels is central for what follows. $A_g$ contains the gradient of the wavefunction, so it grows without bound as the electron localizes, which is what happens as $\alpha$ increases (Fig.~\ref{fig:density}). $A_f$ is a density integrated against a bounded function and cannot grow in the same way.

\begin{figure}[t]
\centering
\includegraphics[width=\linewidth]{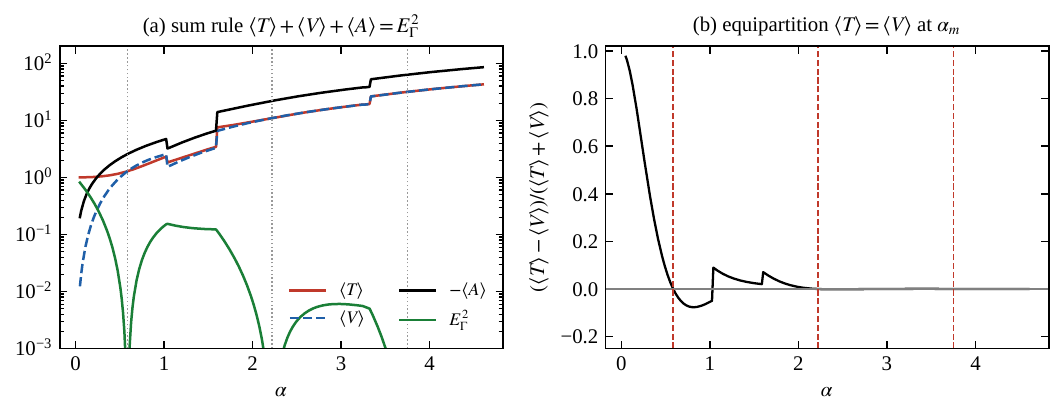}
\caption{Energy balance of the lowest state at $\Gamma$, computed from Eq.~\eqref{eq:sumrule} with a normalized two-layer wavefunction. (a) Kinetic energy $\expval{T}$, confinement energy $\expval{V}$, interlayer energy $-\expval{\mathcal A}$ and $E_\Gamma^2$. (b) Relative difference between kinetic and confinement energy; red dashed lines mark the magic angles. The two energies become equal at the magic angles (exactly in the limit of large $m$, Appendix~\ref{app:equipartition}).}
\label{fig:energy}
\end{figure}

Figure~\ref{fig:energy}(a) shows the three contributions for the lowest state at $\Gamma$. The kinetic and confinement energies grow roughly as $\alpha^2$ and are almost equal for $\alpha\gtrsim2$, and the interlayer term almost cancels them. $E^2_\Gamma$ is the small remainder, and it vanishes at each magic angle. The jumps at $\alpha\approx1.05$, $1.6$ and $3.3$ occur where the lowest state at $\Gamma$ changes character because the flat band hybridizes with the next band (Sec.~\ref{sec:topology}).

\subsection{Why the first magic angle is different}

For small $\alpha$ the zero-order state at $\Gamma$ is a single plane-wave shell with $\expval{T}=1$ and $\expval{V}=0$. First-order perturbation theory (Appendix~\ref{app:pert}) gives $\expval{A_g}=-3\alpha$ and $\expval{A_f}=-\alpha$, so that
\begin{equation}
E^2(\Gamma) \approx 1-4\alpha.
\label{eq:EGammaapprox}
\end{equation}
This agrees with the numerics of Fig.~\ref{fig:energy} for small $\alpha$ (for example $E^2_\Gamma=0.81$ at $\alpha=0.05$). Extrapolated, it would put the first zero at $\alpha\approx1/4$. The exact value is $0.586$, which is expected since Eq.~\eqref{eq:EGammaapprox} is only first order.

At the Dirac point $K$ the energy vanishes for any $\alpha$, and the zero mode has the form $\Psi(\vec r)=(g(\vec r),\ii g(-\vec r))\Phi_K(\vec r)$, with $g$ Bloch-periodic and $\Phi_K$ playing the role of a lowest-Landau-level envelope \cite{wang2021b,ledwith2021}. This gives $\expval{A^\dagger}_{12}=\expval{A}_{21}$ and
\begin{equation}
\expval{\mathcal A} = 2\expval{A}_{21} = -\expval{T}-\expval{V}.
\label{eq:KpointBalance}
\end{equation}
The overlap term is $\expval{A_f}\propto\int g^*(\vec r)g(-\vec r)|\Phi_K(\vec r)|^2\cos(\vec q_\mu\cdot\vec r)$ and is bounded, while $\expval{A_g}$ involves $\nabla\psi_j$. Perturbatively (Appendix~\ref{app:pert}) $\expval{A_g}/\expval{A_f}\approx\alpha^2$. The two channels are therefore comparable only for $\alpha\lesssim1$, which is where $\alpha_1$ sits. For $\alpha\gg1$ the current channel $A_g$ dominates completely, and the physics is controlled by orbital, quantum-Hall-like currents rather than by the overlap between layers. This is the basic reason why the first magic angle differs from all the others \cite{navarro2022}. \begin{figure}[t]
\centering
\includegraphics[width=0.62\linewidth]{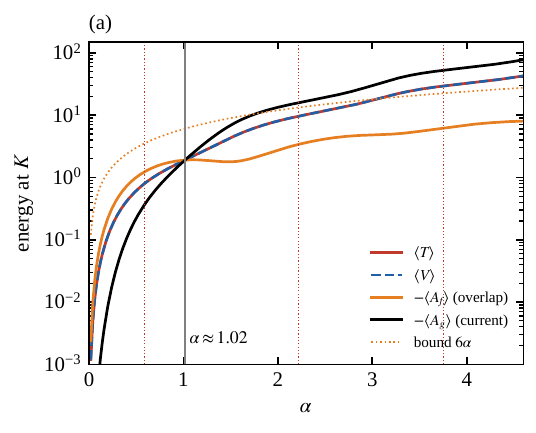}
\caption{Zero mode at the Dirac point $K$ as a function of $\alpha$: kinetic energy $\expval{T}$, confinement energy $\expval{V}$ (identical to $\expval{T}$ for every $\alpha$), and the two interlayer channels, the bounded overlap term $-\expval{A_f}$ and the current term $-\expval{A_g}$. The dotted orange line is the bound $|\expval{A_f}|\le6\alpha$. The two channels cross at $\alpha\approx1.02$ (grey line); dotted red lines mark $\alpha_1$, $\alpha_2$, $\alpha_3$.}
\label{fig:kpoint}
\end{figure}

Figure~\ref{fig:kpoint} shows these channels computed exactly at $K$. At $K$ the kinetic and confinement energies coincide for every $\alpha$ \cite{navarro2022}, and their sum is exactly cancelled by the interlayer term, Eq.~\eqref{eq:KpointBalance}. For small $\alpha$ the numerics reproduce the perturbative results of Appendix~\ref{app:pert}, $\expval{A_f}\approx-6\alpha^2$ and $\expval{A_g}\approx-6\alpha^4$. The overlap channel then saturates well below its bound $6\alpha$, while the current channel keeps growing roughly as $\alpha^2$. The two cross at $\alpha\approx1.02$. At $\alpha_1$ the overlap term still dominates ($\expval{A_f}\approx-1.21$ against $\expval{A_g}\approx-0.35$), while at $\alpha_2$ the current term is already about five times larger and at $\alpha_3$ about eight times larger. The first magic angle is the only one on the overlap-dominated side of the crossing.

Numerically, $\expval{A_g}/\alpha$ organizes into plateaus that change at the hybridization points between magic angles, in the same places where the jumps of Fig.~\ref{fig:energy} appear.

\subsection{The 3/2 rule from a rescaling of the potential}
\label{sec:rescaling}

We now explain Eq.~\eqref{eq:32rule}, following Ref.~\cite{navarro2023a}. Take the zero-mode equation $D(\vec r)(\psi_1,\psi_2)^T=0$ at $\Gamma$ and use Eq.~\eqref{eq:parityrelation} to eliminate $\psi_2$. For the layer-1 component this gives
\begin{equation}
\bar\partial\psi_1(\vec r)=\alpha\,\mu_\alpha\,U(\vec r)\,\psi_1(-\vec r).
\label{eq:zeromodeGamma}
\end{equation}
The equation couples $\vec r$ with $-\vec r$, so it is natural to work with the symmetric and antisymmetric combinations $\psi_\pm$, which satisfy the same equation, $\bar\partial\psi_\pm(\vec r)=\alpha\mu_\alpha U(\vec r)\psi_\pm(-\vec r)$. Now rescale the coordinates,
\begin{equation}
\vec r' = \alpha\,\vec r, \qquad \bar\partial=\alpha\,\bar\partial',
\end{equation}
to obtain
\begin{equation}
\bar\partial'\psi_\pm(\vec r') = \mu_\alpha\, U\!\left(\frac{\vec r'}{\alpha}\right)\psi_\pm(-\vec r').
\label{eq:rescaled}
\end{equation}
The coupling has disappeared from the right-hand side. It now appears only in the argument of the potential. Increasing $\alpha$ is therefore the same as stretching the moir\'e potential by a factor $\alpha$, i.e., solving the same equation on a moir\'e cell that is $\alpha$ times larger. This is the rescaling of the potential.

\begin{figure}[t]
\centering
\includegraphics[width=\linewidth]{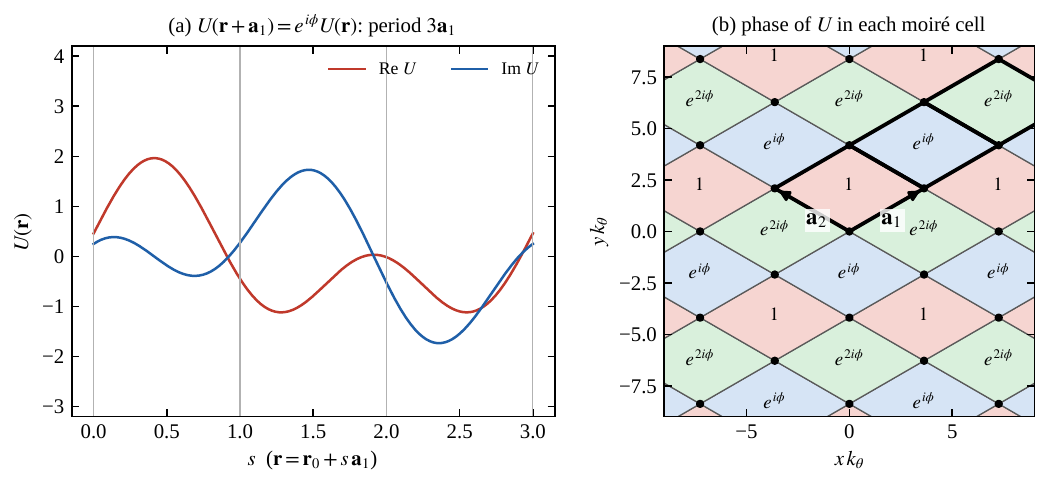}
\caption{Why the magic-angle period is related to the number $3$. (a) Real and imaginary parts of $U(\vec r)$ along a line parallel to $\vec a_1$: the function is not periodic with period $\vec a_1$ but with period $3\vec a_1$. (b) Phase picked up by $U$ in each moir\'e cell, $e^{\ii(n_1+n_2)\phi}$ for the cell at $n_1\vec a_1+n_2\vec a_2$. The pattern repeats on a supercell containing three moir\'e cells, the analogue of the magnetic supercell in the quantum Hall effect.}
\label{fig:rescaling}
\end{figure}

Rescaling alone does not select any value of $\alpha$. What does is the boundary condition. The coupling $U(\vec r)$ is not periodic on the moir\'e cell. A translation by a lattice vector multiplies it by a phase,
\begin{equation}
U(\vec r+\vec a_{1,2}) = e^{\ii\phi}\,U(\vec r), \qquad \phi=\frac{2\pi}{3},
\label{eq:Uphase}
\end{equation}
because $\vec q_\nu\cdot\vec a_{1,2}=-\phi$ modulo $2\pi$ for all three vectors. The potential only comes back to itself after three translations, as shown in Fig.~\ref{fig:rescaling}. This is the same situation as for the magnetic translations of an electron in a magnetic field, where one has to use a magnetic supercell that encloses an integer number of flux quanta. In the rescaled equation~\eqref{eq:rescaled} the lattice vectors become $\vec a'_{1,2}=\alpha\,\vec a_{1,2}$, and the plane waves that make up $U$ pick up phases $e^{-\ii\vec q_\nu\cdot\vec a'_{1,2}}=e^{\ii\alpha\phi}$. If
\begin{equation}
\alpha = 3n, \qquad n=1,2,\dots,
\label{eq:alpha3m}
\end{equation}
these phases are $e^{\ii 3n\phi}=e^{2\pi\ii n}=1$, and the stretched potential satisfies the same boundary conditions as the original one on a larger cell \cite{navarro2023a}. In other words, if we increase $\alpha$ by $3$, we return to an equivalent problem. Solutions with a given parity $\mu_\alpha$ therefore repeat with period $3$ in $\alpha$.

\begin{figure}[t]
\centering
\includegraphics[width=\linewidth]{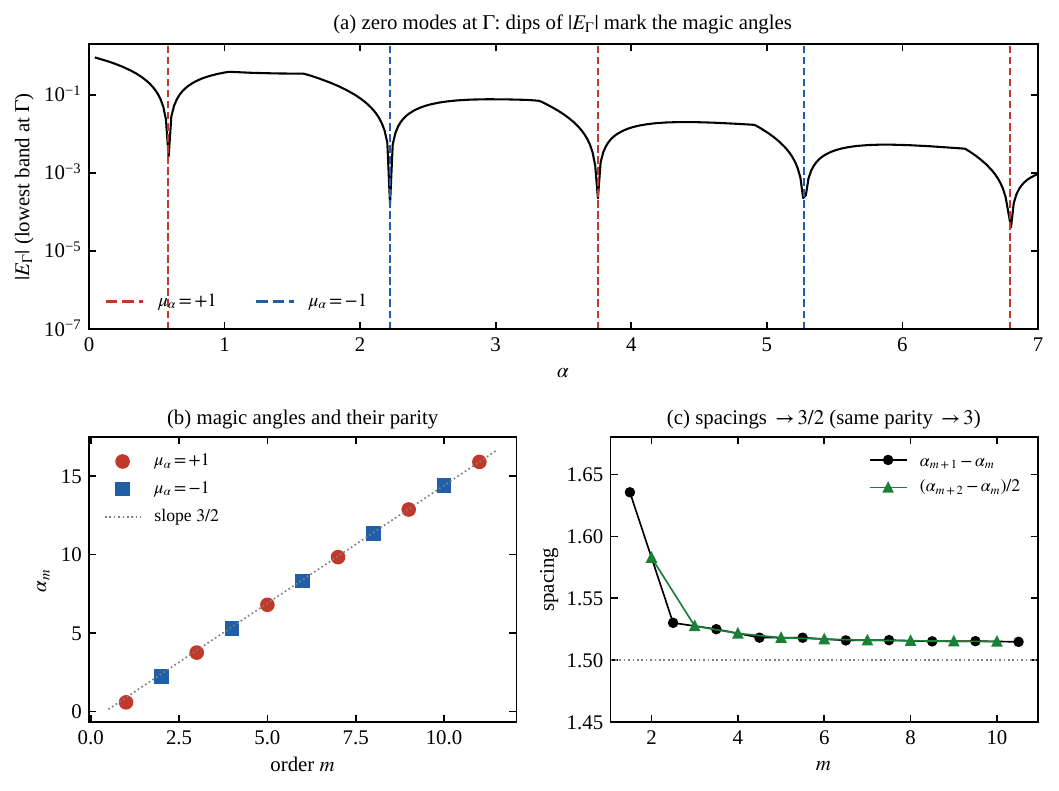}
\caption{(a) Lowest energy at $\Gamma$ versus $\alpha$ (log scale). The dips are the magic angles, colored by the parity of the zero mode ($\mu_\alpha=+1$ red, $-1$ blue); their depth is limited by the $\alpha$ grid. (b) Magic angles versus order $m$: the two parities alternate. (c) Spacing between consecutive magic angles and half the spacing between magic angles of the same parity. Both approach $3/2$ (from above; $1.515$ at $m=9$).}
\label{fig:magic}
\end{figure}

The two parities give two families of magic angles, each with spacing $3$. Numerically they are interleaved, as Fig.~\ref{fig:magic} shows: the parity alternates strictly, $+,-,+,-,\dots$, and each family sits roughly half a period away from the other. Combining the two,
\begin{equation}
\alpha^{(\mu)}_{n+1}-\alpha^{(\mu)}_{n}\to 3 \quad\Longrightarrow\quad \alpha_{m+1}-\alpha_m\to\frac32 .
\label{eq:32derived}
\end{equation}
This is the $3/2$ rule. Three comments are in order.

First, the argument is asymptotic. It assumes that the zero mode is controlled by the local structure of the potential, which is true in the boundary-layer regime $\alpha\gg1$ where the wavefunction is concentrated in small regions (Sec.~\ref{sec:localization}). For the first magic angle the spacing is $1.636$, and the approach to $3/2$ is slow; Table~\ref{tab:magicangles} gives $1.515$ at $m=9$.

Second, the argument fixes the spacing, not the magic angles themselves. The values $\alpha_m$ still have to be computed, and so does the offset between the two parity families.

Third, the whole argument rests on the phase in Eq.~\eqref{eq:Uphase}. If the interlayer potential were periodic on the moir\'e cell, the three-cell supercell would not exist, and nothing would force a period in $\alpha$. We test this directly in Sec.~\ref{sec:abelian}.

The rescaling argument fixes the \emph{spacing} of the magic angles. It does not say what the zero modes look like. That is the role of the harmonic-oscillator (Landau-level) description of Sec.~\ref{sec:LLL}, which is local and gives the shape and width of the zero modes but, by itself, does not select any value of $\alpha$. The two pieces are complementary: the local Landau-level structure gives the states, and the global three-cell boundary condition quantizes $\alpha$.

\subsection{Equipartition and quantized angular momentum}

The rescaling picture has consequences for the energies at the magic angles. Applying $\partial$ to Eq.~\eqref{eq:zeromodeGamma} and using $\partial\bar\partial=\nabla^2$ leads to a second-order equation for $\psi_\pm$ (Appendix~\ref{app:equipartition}). Its expectation value gives the equipartition theorem \cite{navarro2023a}
\begin{equation}
\expval{T}_m = \alpha_m^2\expval{|U(\vec r)|^2}_m,
\label{eq:equipartition}
\end{equation}
i.e., kinetic and confinement energies are equal at the magic angles, as in a harmonic oscillator. This is what Fig.~\ref{fig:energy}(b) shows. For large $m$ the state is concentrated where $|U|^2\approx1$ (Sec.~\ref{sec:localization}), so $\expval{T}_m\approx\alpha_m^2$. Taking the off-diagonal expectation value of $H^2$ at a magic angle gives, with the normalization of Ref.~\cite{navarro2023a},
\begin{equation}
\expval{1|A^\dagger|2}_m + \expval{2|A|1}_m = -\alpha_m .
\label{eq:offdiagquant}
\end{equation}
Writing $A^\dagger$ in polar coordinates around a localization center (Appendix~\ref{app:polar}), the interlayer operator is dominated by the angular momentum $L_\nu=-\ii\partial_\nu$, with a radial residue that becomes small as $m$ grows. Consistency with Eq.~\eqref{eq:offdiagquant} then requires
\begin{equation}
L_\nu\,\psi_\pm(\vec r) \approx m\,\psi_\pm(\vec r),
\label{eq:Lnuquant}
\end{equation}
so the $m$-th zero mode carries about $m$ units of orbital angular momentum around its localization center. Numerically, $\expval{L_z}$ increases by one unit at each magic angle \cite{navarro2023a}. Since $\alpha_m\approx 3m/2$ for large $m$, Eq.~\eqref{eq:offdiagquant} says that the interlayer current grows by $3/2$ from one magic angle to the next: the interlayer currents are quantized along with the angular momentum.

\subsection{Harmonic oscillator, Landau levels and the quantum Hall effect}
\label{sec:LLL}

We now come back to the connection with the quantum Hall effect found in Ref.~\cite{navarro2023a}, and we make it more precise. The key observation is that the coupling $U(\vec r)$ has a simple zero at the AA point. Expanding to first order around $\vec r=0$,
\begin{equation}
U(z,z^*) = \sum_{\mu}(\ii Q_\mu)\,e^{-\ii(Q_\mu^*z+Q_\mu z^*)/2}\approx \frac{3}{2}\,z, \qquad z=x+\ii y,
\label{eq:Ulinear}
\end{equation}
where we used $\sum_\mu Q_\mu=\sum_\mu Q_\mu^2=0$ and $\sum_\mu|Q_\mu|^2=3$. The confinement potential is then harmonic, $\alpha^2|U|^2\approx(3\alpha/2)^2|z|^2$, and the off-diagonal operator becomes $A^\dagger\approx-3\ii L_z-3\ii$ \cite{navarro2023a}. With the parity relation~\eqref{eq:parityrelation}, the zero-mode equation of $H^2$ reduces to
\begin{equation}
\left[4p_zp_z^* + \left(\frac{3\alpha}{2}\right)^2 |z|^2 - 3\alpha L_z\right]\psi(z,z^*) = 3\alpha\,\psi(z,z^*).
\label{eq:Landau}
\end{equation}
This is a two-dimensional harmonic oscillator with an extra $-3\alpha L_z$ term. It is worth recognizing it for what it is: since $4p_zp_z^*=p_x^2+p_y^2$, the left-hand side is exactly $(\vec p+\vec{\mathcal A})^2$ for a particle in a \emph{uniform} magnetic field $B_{\rm eff}$ in the symmetric gauge, $\vec{\mathcal A}=\tfrac{B_{\rm eff}}{2}(-y,x)$, with
\begin{equation}
B_{\rm eff}=3\alpha, \qquad \ell_B=\frac{1}{\sqrt{B_{\rm eff}}}=\frac{1}{\sqrt{3\alpha}} .
\label{eq:Beff}
\end{equation}
The right-hand side, $3\alpha=B_{\rm eff}$, is the energy of the lowest Landau level of $(\vec p+\vec{\mathcal A})^2$. Introducing the cyclotron and guiding-centre ladder operators, $(\vec p+\vec{\mathcal A})^2=B_{\rm eff}(2a^\dagger a+1)$, and Eq.~\eqref{eq:Landau} says $a^\dagger a=0$: the zero modes are lowest-Landau-level states. Their general form is
\begin{equation}
\psi(z,z^*)=f(z)\,e^{-|z|^2/4\ell_B^2},
\label{eq:LLLform}
\end{equation}
with $f(z)$ analytic, which can be checked directly on the first-order equation $\bar\partial\psi_\pm=\pm\mu_\alpha\tfrac{3\alpha}{2}z\,\psi_\pm$ obtained from Eqs.~\eqref{eq:zeromodeGamma} and~\eqref{eq:Ulinear}, since $\bar\partial e^{-|z|^2/4\ell_B^2}=-(z/2\ell_B^2)e^{-|z|^2/4\ell_B^2}$. Only the combination with $\pm\mu_\alpha=-1$ has a decaying Gaussian, so the parity of the magic angle decides whether the zero mode is the symmetric or the antisymmetric one.

This identification clarifies several points.

\emph{The width.} The magnetic length is $\ell_B=1/\sqrt{3\alpha}$, which is exactly the width $\sigma$ of the coherent states of Sec.~\ref{sec:localization}. The coherent Landau states are lowest-Landau-level states with a definite guiding centre $\lambda$, obtained with $f(z)=f_\lambda(z)=e^{\lambda^*z/2\ell_B^2}$, which moves the Gaussian from the origin to $\lambda$ without changing its width.

\emph{The angular momentum.} In the lowest Landau level the cyclotron quantum number vanishes, but the guiding-centre oscillator is free: all states $z^n e^{-|z|^2/4\ell_B^2}$, $n=0,1,2,\dots$, have the same energy and angular momentum $L_z=n$ about the origin. A state whose guiding centre sits at a distance $|\lambda|$ from AA therefore carries an angular momentum about the AA point that grows as $|\lambda|^2/\ell_B^2\propto\alpha$. The oscillator ``ground state'' in Eq.~\eqref{eq:Landau} does not force $L_z=0$, and the growing angular momentum found numerically in Ref.~\cite{navarro2023a} is the angular momentum of the guiding centre, not of a cyclotron orbit.

\begin{figure}[t]
\centering
\includegraphics[width=\linewidth]{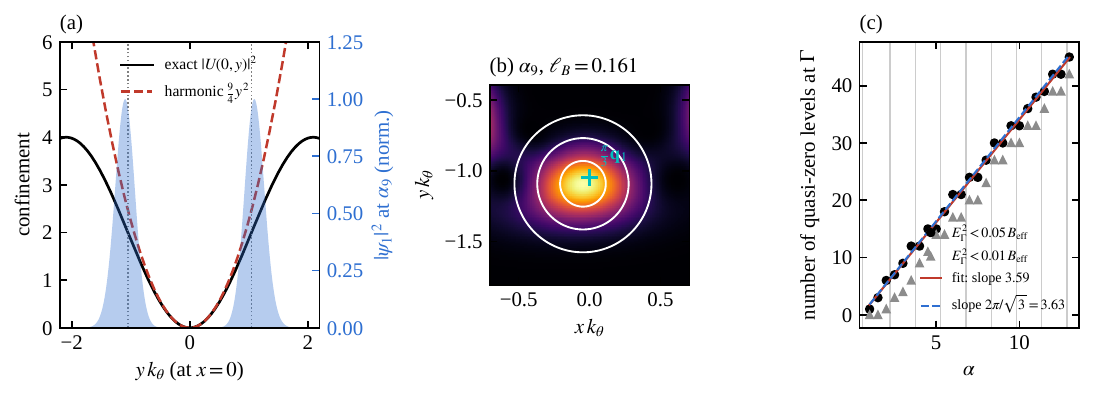}
\caption{Harmonic-oscillator (Landau-level) picture of the zero modes. (a) Exact confinement $|U(0,y)|^2$ along the line through AA and $\vec q_1$, compared with the harmonic approximation $(9/4)y^2$ of Eq.~\eqref{eq:Ulinear}; the shaded curve is the layer-1 density of the $\Gamma$ zero mode at $\alpha_9$ (right axis), located at $|y|\approx1.1$, far from the origin where the expansion is made. (b) Numerical density of the $\Gamma$ zero mode at $\alpha_9$ around the localization centre near $(\pi/3)\vec q_1$ (cross) (colour) and contours of the lowest-Landau-level coherent state $|f_\lambda(z)e^{-|z|^2/4\ell_B^2}|^2$ with $\ell_B=1/\sqrt{3\alpha_9}$ centred at the numerical maximum (white lines, at $e^{-1/2}$, $e^{-2}$ and $e^{-9/2}$ of the maximum, i.e., $1\ell_B$, $2\ell_B$, $3\ell_B$). (c) Number of quasi-zero levels at $\Gamma$, Eq.~\eqref{eq:quasizero}: bands whose energy at $\Gamma$ satisfies $E^2<\epsilon B_{\rm eff}$, for $\epsilon=0.05$ (dots) and $0.01$ (triangles), computed from the singular values of $D$ at $\Gamma$. Solid line: linear fit, slope $3.59$; dashed line: slope $2\pi/\sqrt3\approx3.63$, half the number of flux quanta of $B_{\rm eff}$ per moir\'e cell per unit $\alpha$. The linear growth mirrors the field-proportional degeneracy of a Landau level. Vertical lines: magic angles.}
\label{fig:LLL}
\end{figure}

\emph{The problem at $\vec r=0$.} Equation~\eqref{eq:Ulinear} is a Taylor expansion about the AA point and is strictly valid only for $|\vec r|\ll1$, whereas at high-order magic angles the electrons sit at $|\vec r|\approx\pi/3$, away from the origin (Fig.~\ref{fig:density}). The oscillator of Eq.~\eqref{eq:Landau}, read literally, is centred at $\vec r=0$ where there is no charge. The correct reading is the one above: the expansion about the zero of $U$ fixes the effective field $B_{\rm eff}$ and hence $\ell_B$, while the states themselves are Landau-level states centred on the localization centres $\lambda\approx\pm(\pi/3)\vec q_\mu$. Around each centre the Gaussian of Eq.~\eqref{eq:LLLform} is the ground state of a harmonic oscillator of the same frequency, displaced to $\lambda$ (a magnetic translation does not change the energy). Two facts make this work beyond $|\vec r|\ll1$. First, along the directions of the $\vec q_\mu$ the linear form is still reasonable at $|\vec r|=\pi/3$: $|U|^2=2$ there, against $(9/4)(\pi/3)^2=2.47$ from Eq.~\eqref{eq:Ulinear} (Fig.~\ref{fig:LLL}a). Second, for $\alpha\gg1$ the width $\ell_B\to0$, so each Gaussian only probes a small neighbourhood of its centre. The picture fails for the first magic angle, where $\ell_B=0.75$ is comparable with the distance between centres and the Gaussians overlap. Appendix~\ref{app:HOlimit} discusses these limits in more detail.

\emph{Relation to the quantum Hall effect and to the 3/2 rule.} In a uniform field, the lowest Landau level has a degeneracy proportional to the field: one state per flux quantum threading the sample. If the zero modes of TBG are lowest-Landau-level states in the field $B_{\rm eff}=3\alpha$, the number of states that are (almost) at zero energy should grow linearly with $\alpha$. We test this with the quantity shown in Fig.~\ref{fig:LLL}(c), which we call the number of quasi-zero levels. It is computed as follows. At the single momentum $\vec k=\Gamma$ we build the matrix of $D(\vec r)$ in the plane-wave basis of Appendix~\ref{app:bloch} and obtain all its singular values $s_n$; these are the energies $|E_n(\Gamma)|$ of all the bands of $H$ at $\Gamma$ (each $\pm$ pair counted once). We then count how many of them satisfy
\begin{equation}
E_n^2(\Gamma)=s_n^2<\epsilon\,B_{\rm eff}=3\epsilon\,\alpha, \qquad \epsilon=0.05,
\label{eq:quasizero}
\end{equation}
i.e., how many bands lie far below the Landau energy scale $B_{\rm eff}$ at $\Gamma$. At a magic angle two of them are the exact flat bands; the others are remote bands that, in the chiral model, are squeezed exponentially towards $E=0$ as $\alpha$ grows \cite{hitrik2023}, and which in the Landau-level picture belong to the same lowest-level manifold. The count grows in steps of three, reflecting the $C_3$ multiplets at $\Gamma$, and is well described by a straight line. A least-squares fit over $1\le\alpha\le13$ gives
\begin{equation}
N_{\rm qz}(\alpha)\approx(3.59\pm0.04)\,\alpha-1.8 .
\label{eq:Nqz}
\end{equation}
The slope does not depend on the precise threshold: $\epsilon=0.01$ gives $3.63$, while a much looser $\epsilon=0.2$, which starts to include the tails of the next manifold, gives $3.84$. The slope has a simple interpretation. The number of flux quanta of $B_{\rm eff}$ through one moir\'e cell of area $A_M=8\pi^2/3\sqrt3$ is $B_{\rm eff}A_M/2\pi=(4\pi/\sqrt3)\,\alpha\approx7.26\,\alpha$, and half of it is $(2\pi/\sqrt3)\,\alpha\approx3.63\,\alpha$, which agrees with the fitted slope within its uncertainty (dashed line in Fig.~\ref{fig:LLL}c). A natural origin for the factor $1/2$ is that, as shown above, only one of the two parity sectors $\psi_\pm$ supports decaying (normalizable) lowest-Landau-level states. We regard this agreement as suggestive rather than as a derivation, but it gives a quantitative meaning to the statement that the chiral zero-mode manifold behaves like a Landau level whose degeneracy is set by $B_{\rm eff}=3\alpha$. In units of the magic-angle spacing, about $3.6\times3/2\approx5.4$ new quasi-zero levels enter at $\Gamma$ between two consecutive magic angles. The quantization of the magic angles itself, however, does not come from Eq.~\eqref{eq:Landau}, which has lowest-Landau-level solutions for every $\alpha$. It comes from the moir\'e boundary condition of Sec.~\ref{sec:rescaling}: the three-cell supercell plays the role of the magnetic unit cell, and only when the rescaled potential is compatible with it, $\Delta\alpha=3$ within each parity family, is a zero mode compatible with Bloch periodicity. This is the sense in which the magic angles are a quantum Hall-like quantization: a local Landau-level structure with $B_{\rm eff}=3\alpha$, plus a commensurability condition between the effective magnetic length and the moir\'e lattice.

\section{Coherent States, Confinement and Non-Abelian Magnetism}
\label{sec:localization}

\subsection{Coherent Landau states}

\begin{figure}[t]
\centering
\includegraphics[width=\linewidth]{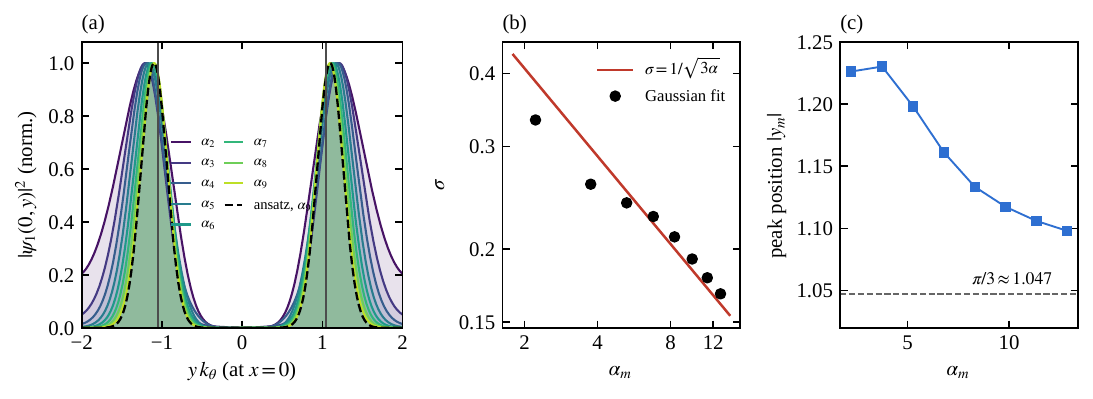}
\caption{Convergence of the zero modes to coherent Landau states. (a) Layer-1 density of the $\Gamma$ zero mode along $x=0$ at the magic angles $\alpha_2$ to $\alpha_9$, each normalized to its maximum, compared with the coherent-state ansatz of Eq.~\eqref{eq:appcoherentfinal} at $\alpha_9$ (dashed; Gaussians of width $\sigma=1/\sqrt{3\alpha_9}$ at the numerical peak positions). Vertical lines: $y=\pm\pi/3$, the asymptotic position of the localization centres. (b) Width of the peak from a Gaussian fit (dots) compared with $\sigma=1/\sqrt{3\alpha}$ (line), without fitted parameters; the relative difference is below $5\%$ from $\alpha_5$ on. (c) Position of the density maximum, which approaches $\pi/3\approx1.047$ (dashed line) as $m$ grows.}
\label{fig:coherent}
\end{figure}

Figure~\ref{fig:density} suggests that for large $m$ the zero mode at $\Gamma$ becomes a set of Gaussian spots centred near $\pm(\pi/3)\vec q_\nu$. This can be made quantitative \cite{navarro2023a}. In the limit $\alpha\to\infty$ we use the boundary-layer idea: the wavefunction is negligible except where $\bar\partial\psi$ is of the same order as $\alpha U\psi$. Expanding $U$ to first order around the centres $Q_\mu=q_{\mu,x}+\ii q_{\mu,y}$ and trying a coherent-state ansatz
\begin{equation}
\psi_\pm(z,z^*)\propto\sum_\mu\big(f_{Q_\mu}(z)\pm f_{-Q_\mu}(z)\big)e^{-|z|^2/4\sigma^2},
\end{equation}
with $f_\lambda(z)$ the standard coherent-state envelope (Appendix~\ref{app:pert}), fixes the width self-consistently,
\begin{equation}
\sigma = \frac{1}{\sqrt{3\alpha}}.
\label{eq:sigmalimit}
\end{equation}
The density near each centre is then a Gaussian,
\begin{equation}
\rho(\vec r)\approx\frac{1}{2\pi\sigma^2}\,e^{-|\vec r-\vec R|^2/2\sigma^2},
\end{equation}
with $|\vec R|$ converging to $\pi/3\approx1.05$ (numerically, along the $y$ axis the maxima sit at $|\vec R|=1.19$, $1.13$ and $1.10$ for $\alpha_4$, $\alpha_6$ and $\alpha_8$, Fig.~\ref{fig:localization}d). Small deviations from a pure Gaussian come from a Weierstrass-sigma factor known from exact constructions \cite{wang2021a,popov2021}. The prediction $\sigma=1/\sqrt{3\alpha}$ has no free parameter and agrees well with Gaussian fits of the numerical zero modes for $m\ge3$ \cite{navarro2023a}; Fig.~\ref{fig:localization}(e) shows the comparison at $\alpha_8$, and Fig.~\ref{fig:coherent} shows it systematically for $\alpha_2$ to $\alpha_9$.

A Gaussian of width $\sigma$ in real space has width $1/(2\sigma)=\sqrt{3\alpha}/2$ in momentum space, so each spot saturates the uncertainty relation, $\Delta x\,\Delta k_x=1/2$. The high-order zero modes are therefore minimal-uncertainty coherent states, like the states of the lowest Landau level. Their form in reciprocal space is again Gaussian, which is the self-duality discussed in Ref.~\cite{navarro2024}. It is worth contrasting this with the exact zero mode at small $\alpha$, a ratio of theta functions with a complicated node structure. The simple form only emerges asymptotically.

Figure~\ref{fig:coherent} tests these statements up to the ninth magic angle. The density peaks become Gaussian and narrow as $\alpha$ grows, and the fitted widths follow $\sigma=1/\sqrt{3\alpha}$; the ratio $\sigma_{\rm fit}\sqrt{3\alpha}$ goes from $0.95$ at $\alpha_2$ through $0.86$ at $\alpha_3$ to $1.04$--$1.05$ for $\alpha_6$--$\alpha_9$. The peak position moves from $1.23$ at $\alpha_2$ to $1.10$ at $\alpha_9$, slowly approaching the asymptotic value $\pi/3\approx1.047$: a fit $|y_m|=r_\infty+c/\alpha_m$ to $\alpha_4,\dots,\alpha_9$ gives $r_\infty\approx1.03$, within $2\%$ of $\pi/3$. At $\alpha_9$ the ansatz and the numerical profile are almost indistinguishable. The asymptotic radius $\pi/3$ has a simple meaning. Along the line from AA towards $\vec q_\mu$ the phases of the moir\'e harmonics are $\vec b_\nu\cdot\vec r=3r/2$, so at $r=\pi/3$ they equal $\pi/2$: the periodic confinement $2-2\cos(3r/2)$ crosses its mean value there and its slope, i.e., the restoring force that balances the centrifugal barrier of the next subsection, is maximal. We identify $\pi/3$ with the minimum of the periodic effective confinement potential felt by the zero modes; for large $\alpha$ the probability condenses at $\pm(\pi/3)\vec q_\mu$.

\subsection{Where the electrons sit}

\begin{figure}[!t]
\centering
\includegraphics[width=\linewidth]{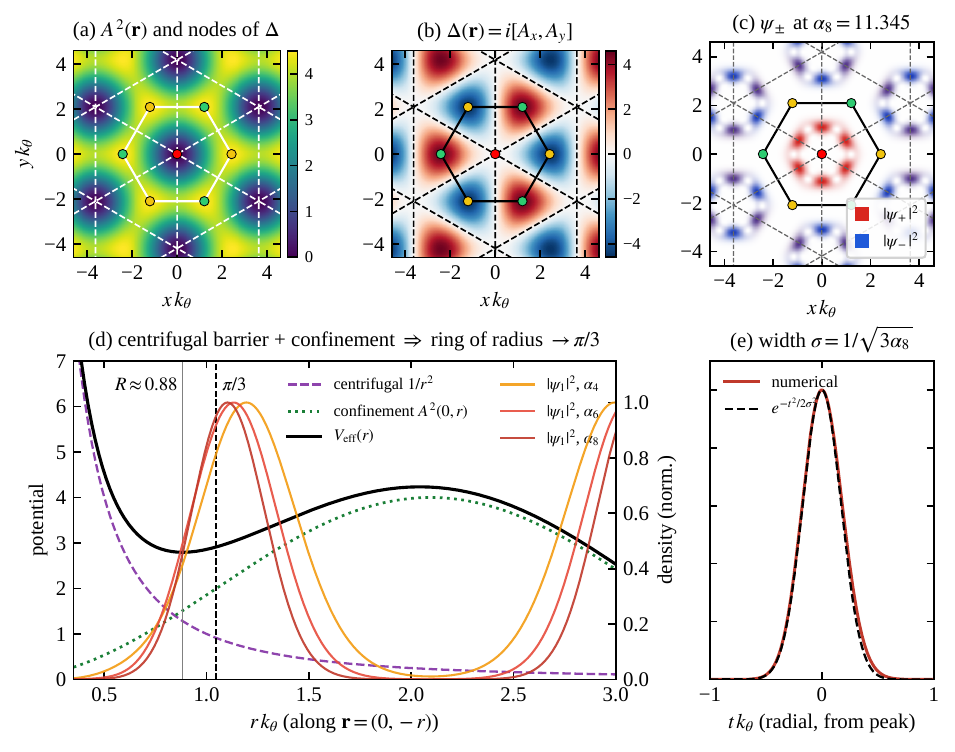}
\caption{Potentials and localization of the zero modes. (a) Symmetric confinement $A^2(\vec r)$ of Eq.~\eqref{eq:A2Delta}; dashed lines are the nodal lines of $\Delta(\vec r)$. (b) Antisymmetric commutator $\Delta(\vec r)=\ii[A_x,A_y]$ with its nodal lines. (c) Densities of the parity combinations $\psi_\pm=\psi_1(\vec r)\pm\psi_1(-\vec r)$ of the $\Gamma$-point zero mode at $\alpha_8$, each normalized to its maximum: the two sectors occupy different regions of the cell and sit on the nodal lines of $\Delta$. (d) Radial effective potential of Eq.~\eqref{eq:radialeq} along $\vec r=(0,-r)$, i.e., towards $\vec q_1$: the centrifugal barrier and the confinement produce a minimum at $R\approx0.88$. The layer-1 densities at $\alpha_4$, $\alpha_6$, $\alpha_8$ (right axis) peak at $r=1.19$, $1.13$, $1.10$, moving towards $\pi/3\approx1.05$ (dashed line). (e) Radial cut of the density through its maximum at $\alpha_8$ compared with a Gaussian of width $\sigma=1/\sqrt{3\alpha_8}$, with no fitted parameter. Hexagon and dots as in Fig.~\ref{fig:density}.}
\label{fig:localization}
\end{figure}

To see where the localization centres come from, write $\psi_\pm=\psi_1(\vec r)\pm\psi_1(-\vec r)$ and split the gauge field as $\vec A_\pm=(\vec A_+\pm\vec A_-)/2$. The zero-mode equations for $\psi_1(\vec r)$ and $\psi_1(-\vec r)$ combine into \cite{navarro2024}
\begin{equation}
\big(-\nabla^2+\alpha^2 A^2 - \ii\mu_\alpha\alpha(-2\ii\vec A_\mp\cdot\nabla+\nabla\times\vec A_\mp)\big)\psi_\pm
+ \big(\alpha^2\Delta - \ii\mu_\alpha\alpha(-2\ii\vec A_\pm\cdot\nabla+\nabla\times\vec A_\pm)\big)\psi_\mp = 0,
\label{eq:coupledpm}
\end{equation}
where the second term mixes the two parity sectors through $\Delta(\vec r)$. The two potentials involved are shown in Fig.~\ref{fig:localization}(a,b). Numerically, for large $\alpha$ the two sectors occupy different regions of the cell (Fig.~\ref{fig:localization}c), so the mixing term acts on a vanishing overlap and can be dropped. For $\alpha\to\infty$ the curl terms are subleading with respect to the $\alpha^2$ confinement, and using that $A^2$ is even and $\Delta$ odd under $\vec r\to-\vec r$ one arrives at the two conditions
\begin{equation}
\big(-\nabla^2+\alpha^2 A^2(\vec r)\big)\psi_\pm \approx 0, \qquad \Delta(\vec r)\,\psi_\pm \approx 0.
\label{eq:finaldecoupled}
\end{equation}
The second one says that the wavefunction lives near the nodal lines of $\Delta(\vec r)$, where the commutator of Eq.~\eqref{eq:commutatorexplicit} vanishes and the field is locally Abelian. The nodal set consists of three families of straight lines related by $C_3$, some of which pass through the AA points. Figure~\ref{fig:localization}(c) shows that the six maxima of $\psi_+$ lie on the lines through the central AA point, at a distance close to $\pi/3$ from it, while $\psi_-$ occupies a complementary set of segments, consistent with the parity relation~\eqref{eq:parityrelation}.

Along one of these directions, dropping the angular derivatives and using $L_\nu\psi_\pm\approx m\psi_\pm$ with $\alpha_m\approx3m/2$, one obtains a radial equation with a centrifugal barrier,
\begin{equation}
-\frac{\partial^2\psi_\pm}{\partial r^2} + \Big(\frac{9}{4}m^2\frac{1}{r^2}+A^2(\vec r)\Big)\psi_\pm \approx 0.
\label{eq:radialeq}
\end{equation}
In rescaled units the effective potential along the symmetric direction is $V_{\rm eff}(r)=1/r^2+A^2(0,r)$, with $A^2(0,r)=2-2\cos(3r/2)$. Its minimum, from $-2/R^3+3\sin(3R/2)=0$, is at $R\approx0.88$, to be compared with the numerical peak positions $1.10$--$1.19$ (Fig.~\ref{fig:localization}d). The curvature there is
\begin{equation}
\omega^2(R) = \left.\frac{\dd^2V_{\rm eff}}{\dd r^2}\right|_{r=R} = \frac{6}{R^4}+\frac92\cos\!\Big(\frac{3R}{2}\Big) \approx 11.0 .
\label{eq:omega2R}
\end{equation}
The zero mode is therefore close to the ground state of a harmonic well centred at $R$, which explains the Gaussian profile of the previous subsection. This one-dimensional estimate is rough: it neglects the angular structure and the overlap between neighbouring spots, which explains the differences of $20$--$25\%$ with the numerical values. The Gaussian shape itself, however, is reproduced very well (Fig.~\ref{fig:localization}e). As $\alpha$ grows, the dominant Fourier components move to higher reciprocal vectors; Appendix~\ref{app:mathieu} gives a simple analogy with a parametric (Mathieu) oscillator.

\subsection{A non-Abelian quantum Hall problem}

\begin{figure}[t]
\centering
\includegraphics[width=\linewidth]{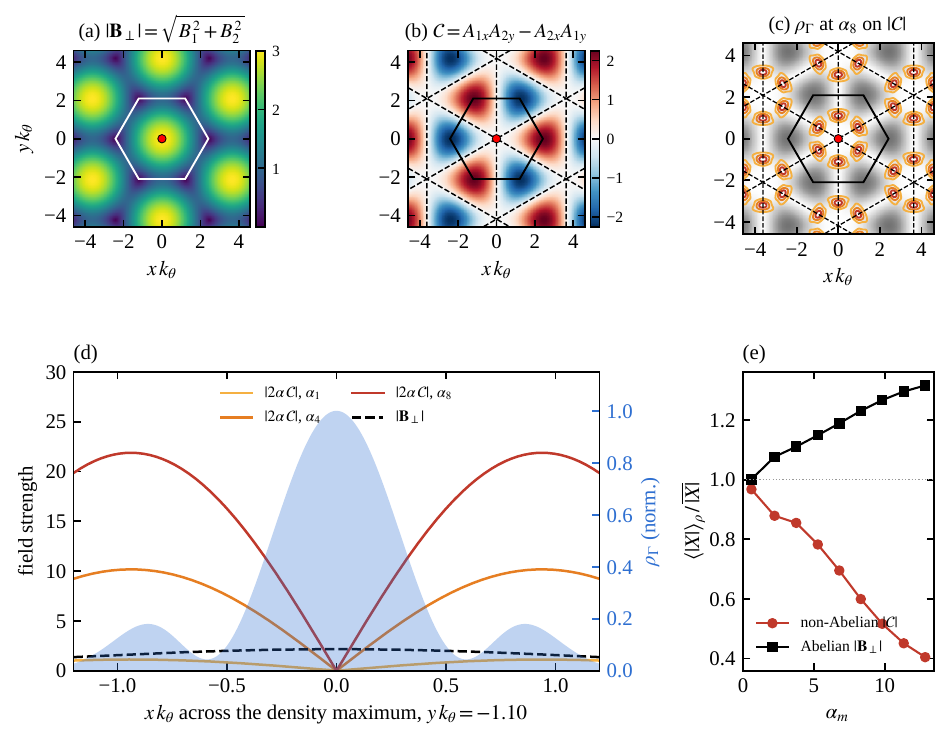}
\caption{Non-Abelian field strength and localization. (a) Magnitude of the Abelian (curl) part $|\vec B_\perp|=\sqrt{B_1^2+B_2^2}$ of $\hat F_{xy}$, Eq.~\eqref{eq:Fsplit}. (b) Non-Abelian commutator part $\mathcal C$ (coefficient of $2\alpha\tau_3$); dashed: nodal lines $\mathcal C=0$, where the field is locally Abelian. (c) Density of the $\Gamma$ zero mode at $\alpha_8$ (contours at $0.2$, $0.5$, $0.8$ of the maximum) on top of $|\mathcal C|$ (grey): the charge sits on the nodal lines. (d) Cut along $x$ through a density maximum ($y=-1.10$): $|2\alpha\,\mathcal C|$ at $\alpha_1$, $\alpha_4$, $\alpha_8$ (solid), $|\vec B_\perp|$ (dashed) and the density at $\alpha_8$ (shaded, right axis). (e) Density-weighted field divided by its cell average at the magic angles $\alpha_1$--$\alpha_9$, for the non-Abelian part $|\mathcal C|$ and the Abelian part $|\vec B_\perp|$.}
\label{fig:Fxy}
\end{figure}

It is instructive to look directly at the field strength that the electron feels. With the matrix potential of Eq.~\eqref{eq:SU2A}, the field-strength operator of Eq.~\eqref{eq:H2enlightening} splits into two pieces with different character \cite{navarrosusy2025},
\begin{equation}
\hat F_{xy}=B_1\tau_1+B_2\tau_2+2\alpha\,\mathcal C\,\tau_3, \qquad
B_j=\partial_xA_{j,y}-\partial_yA_{j,x}, \qquad \mathcal C=A_{1,x}A_{2,y}-A_{2,x}A_{1,y}.
\label{eq:Fsplit}
\end{equation}
The in-plane components $B_{1,2}$ are ordinary curls, as for an Abelian field, and do not depend on $\alpha$. The $\tau_3$ component comes entirely from the commutator $-\ii\alpha[\hat A_x,\hat A_y]$ of Eq.~\eqref{eq:commutatorexplicit}; it is the genuinely non-Abelian part of the field, and it grows linearly with $\alpha$. Figure~\ref{fig:Fxy} shows both pieces and compares them with the zero modes. The Abelian part $|\vec B_\perp|$ is largest at AA and varies smoothly, while $\mathcal C$ changes sign across a set of straight nodal lines related by $C_3$. At high order the zero-mode density sits on these nodal lines, at the points $\approx\pm(\pi/3)\vec q_\mu$ (panel c). A cut across one of the density maxima (panel d) shows the mechanism: the non-Abelian term $|2\alpha\,\mathcal C|$ vanishes linearly on the nodal line and grows away from it with a slope proportional to $\alpha$, forming a V-shaped transverse confinement that becomes steeper at each magic angle, while the Abelian part is almost constant on the scale of the wavefunction. The electron therefore localizes where the field is locally Abelian, as anticipated in Sec.~\ref{sec:localization} from the condition $\Delta(\vec r)\psi_\pm\approx0$.

This can be quantified by weighting each field with the zero-mode density at the magic angles, $\langle|X|\rangle_\rho=\int\rho|X|/\int\rho$, and comparing it with its cell average $\overline{|X|}$ (panel e). For the non-Abelian part the ratio falls from $0.97$ at $\alpha_1$ to $0.40$ at $\alpha_9$: the electron avoids the regions of strong non-commutativity, increasingly so at higher order. For the Abelian part the ratio instead grows from $1.00$ to $1.32$: the electron sits where the ordinary, commuting part of the field is stronger than average. The first magic angle is, once more, the exception, since there the density samples both parts of the field almost uniformly. The high-order zero modes thus behave as Landau-level states of an effectively Abelian field (Sec.~\ref{sec:LLL}) precisely because they are expelled from the non-Abelian regions.

Rescaling $\vec r\to\vec r/\alpha$ in Eq.~\eqref{eq:H2enlightening}, as in Sec.~\ref{sec:rescaling}, gives $H_s^2\equiv(H/\alpha)^2$ \cite{navarro2024},
\begin{equation}
H_s^2 = \big(-\nabla^2+A^2(\vec r/\alpha)\big)\tau_0 + \ii[A_x(\vec r/\alpha),A_y(\vec r/\alpha)]\tau_z - 2\ii\hat A(\vec r/\alpha)\cdot\nabla - \frac1\alpha\hat B(\vec r/\alpha)\cdot\hat\tau,
\label{eq:Hsrescaled}
\end{equation}
where $\hat\tau=(\tau_x,\tau_y)$ acts on the layer pseudospin and $\hat B$ is defined through $F_{xy}=-\hat B\cdot\hat\tau+\ii\alpha[\hat A_x,\hat A_y]$. For a strongly confined state, $\vec A\cdot\vec p$ can be replaced by $-\vec B\cdot\vec L$, so that
\begin{equation}
H_s^2 = \big(-\nabla^2+A^2(\vec r/\alpha)\big)\tau_0 + \ii[A_x(\vec r/\alpha),A_y(\vec r/\alpha)]\tau_z - \hat B(\vec r/\alpha)\cdot\Big(2\hat L+\frac{\hat\tau}{\alpha}\Big).
\label{eq:H2offdiag}
\end{equation}
Only the last term depends explicitly on $\alpha$. It contains the pseudo-Zeeman energy $E_{\rm Zeeman}=\expval{\hat B\cdot\hat\tau}/\alpha$, which decreases as $1/\alpha$, and the orbital magnetic energy $E_{\rm magnetic}=-2\expval{\hat B\cdot\hat L}$, which does not. Numerically the Zeeman term is already about ten times smaller than the orbital one at the third magic angle \cite{navarro2024}. Neglecting it,
\begin{equation}
H_s^2 \;\longrightarrow\; \big(\vec p+\hat A(\vec r/\alpha)\big)^2 + \ii\big[A_x(\vec r/\alpha),A_y(\vec r/\alpha)\big]\tau_z,
\label{eq:nonabelianQHE}
\end{equation}
which is a charged particle in a non-Abelian $SU(2)$ magnetic field with $C_3$ symmetry. At $\Gamma$, using equipartition and $\expval{\Gamma|\Delta|\Gamma}=0$, one finds after normalizing over the three localization centres \cite{navarro2024}
\begin{equation}
\frac{1}{\alpha A_M}\expval{\psi_1(\vec r)|\vec B_\nu(\vec r/\alpha)\cdot\vec L_\nu|\psi_1(-\vec r)} = \frac{\mu_\alpha}{3},
\label{eq:chiralitycontribution}
\end{equation}
with $A_M=8\pi^2/3\sqrt3$ the area of the moir\'e cell. The sign of the orbital magnetic energy is set by the parity, so the circulation reverses between consecutive magic angles. Since the parity alternates (Table~\ref{tab:magicangles}),
\begin{equation}
\mu_{\alpha_{m+1}} = -\,\mu_{\alpha_m},
\label{eq:parityalternation}
\end{equation}
the circulation changes at each hybridization point between magic angles, where the central band exchanges character with the remote bands. This is the same alternation that interleaves the two parity families in Sec.~\ref{sec:rescaling}.

\subsection{Switching off the non-Abelian character}
\label{sec:abelian}

If the three-cell structure of the coupling is responsible for the $3/2$ rule, removing it should destroy the rule. One way to test this is to interpolate between the chiral coupling and an Abelian one,
\begin{equation}
U_\beta(\vec r) = U(\vec r) + \beta\, U(-\vec r), \qquad \beta\in[0,1],
\label{eq:Ubeta}
\end{equation}
keeping the chiral block structure,
\begin{equation}
H_\beta = \begin{pmatrix} 0 & D_\beta^\dagger\\ D_\beta & 0\end{pmatrix},
\qquad
D_\beta(\vec r) = \begin{pmatrix} -\ii\bar\partial & \alpha U_\beta(\vec r) \\ \alpha U_\beta(-\vec r) & -\ii\bar\partial \end{pmatrix}.
\label{eq:Hbeta}
\end{equation}
Squaring works as before for any $\beta$; only the structure of the coupling changes. At $\beta=1$ we have $U_1(-\vec r)=U_1(\vec r)$ and
\begin{equation}
D_1(\vec r) = -\ii\bar\partial\,\tau_0 + \alpha U_1(\vec r)\,\tau_x, \qquad U_1(\vec r) = 2\sum_{\nu=1}^3 e^{\ii(\nu-1)\phi}\cos(\vec q_\nu\cdot\vec r).
\label{eq:D1beta}
\end{equation}
The only matrix left in layer space is $\tau_x$, so the gauge field is Abelian. Moreover $U(-\vec r)$ picks up the opposite phase, $U(-\vec r-\vec a_{1,2})=e^{-\ii\phi}U(-\vec r)$, so for $\beta\neq0$ the coupling no longer has a single well-defined phase under translations, and the argument of Sec.~\ref{sec:rescaling} no longer applies.

A word on how we detect magic angles in this setting. For $\beta\neq0$ the term $U(-\vec r)$ couples a Bloch state at $\vec k$ with states at $\vec k+\vec q_\nu$, which belong to a different coset of the reciprocal lattice. The plane-wave basis must then contain the whole lattice generated by the $\vec q_\nu$, the effective Brillouin zone becomes three times smaller, and $\Gamma$, $K$ and $K'$ fold onto the same momentum ($\Gamma=K+\vec q_1$, $K'=K-\vec q_1$). At $\Gamma$ one therefore always finds the two Dirac zero modes coming from $K$ and $K'$, with $E=0$ for every $\alpha$, and the lowest level at $\Gamma$ says nothing about flat bands. The $M$ point does not fold onto $K$ or $K'$, so its lowest level vanishes only when the band is flat, which is why we use it as the main diagnostic. At $\Gamma$ we plot instead the lowest level above the two folded Dirac modes. For $\beta=0$ this is exactly the quantity shown in Fig.~\ref{fig:magic}(a).

\begin{figure}[t]
\centering
\includegraphics[width=\linewidth]{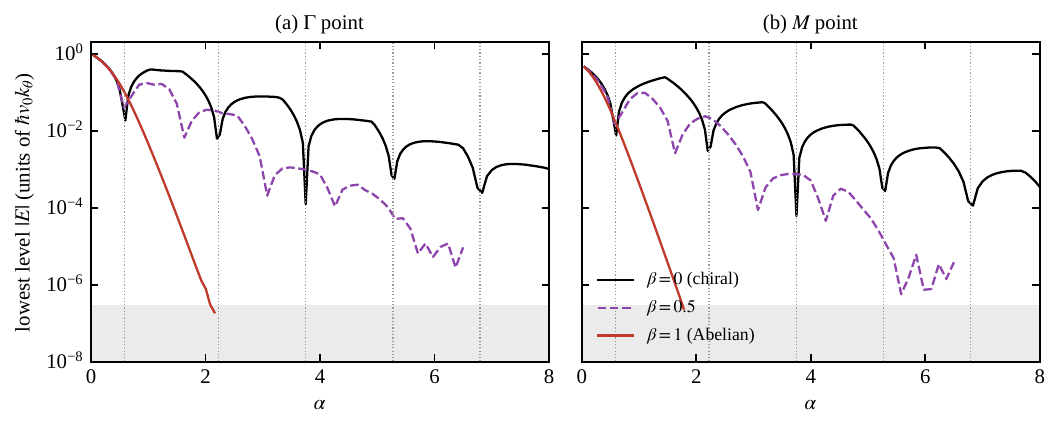}
\caption{Lowest level for the interpolated coupling~\eqref{eq:Ubeta} at (a) $\Gamma$ and (b) $M$. At $\Gamma$ we plot the lowest level above the two Dirac modes of $K$ and $K'$ that fold onto $\Gamma$ for $\beta\neq0$ (see text). Dotted lines: magic angles of the chiral model. For $\beta=0$ the dips at both points coincide with the magic angles. For $\beta=0.5$ the dips move, become shallower and irregularly spaced. For the Abelian limit $\beta=1$ there are no dips: the level decreases exponentially with $\alpha$ until it reaches our numerical precision (grey band, $\sim10^{-7}$). The $\beta=0.5$ curve is shown up to $\alpha=6.5$; beyond $\alpha\approx5$ its small oscillations are limited by the size of the plane-wave basis.}
\label{fig:beta}
\end{figure}

Figure~\ref{fig:beta} shows the result at both points. The two diagnostics agree. For $\beta=0$ they reproduce the magic angles of Table~\ref{tab:magicangles}. For $\beta=0.5$ the dips appear at $\alpha\approx1.63$, $3.08$ and $4.26$, no longer at the chiral magic angles and without a regular spacing, and they are much shallower. At $\beta=1$ the sequence is gone: the lowest level decreases exponentially and smoothly with $\alpha$, at both $\Gamma$ and $M$, without any discrete set of magic angles (see also Ref.~\cite{hitrik2023} for the exponential squeezing of the chiral bands). A $U(1)$ coupling of the same strength does not produce the magic-angle sequence. This supports the view that the non-Abelian structure of the interlayer coupling, and not only its strength, is what fixes the sequence.

\section{Topology Between Magic Angles}
\label{sec:topology}

Between two consecutive magic angles the central bands are no longer flat, and they can touch the remote bands. These touchings change the Chern numbers of the central bands \cite{navarro2026a}. In this section we summarize the resulting phase diagram.

\subsection{Band inversions}

When $\alpha$ is moved away from $\alpha_m$, the central bands disperse, and between $\alpha_m$ and $\alpha_{m+1}$ the gap to the first remote band can close at $\Gamma$ (quadratic touching) or at $M$ (linear touching) \cite{navarro2026a}. In the first window the sequence is $\Gamma\to M\to\Gamma$, at $\alpha\approx1.03$, $1.50$ and $1.65$ ($\theta\approx0.62^\circ$, $0.42^\circ$ and $0.39^\circ$). These are the kinks seen in Figs.~\ref{fig:energy} and~\ref{fig:magic}(a). At higher orders the sequence becomes $M\to\Gamma\to\Gamma$, and once the gap at $\Gamma$ has closed it stays very small over a wide range of $\alpha$.

The inversions reorganize the charge density. Across a $\Gamma$-point inversion, the central band, whose density was concentrated at the boundary between AA and AB/BA regions, exchanges character with the second remote band, which was concentrated at AA; the first remote band is almost unaffected. Across an $M$-point inversion the central and first remote bands exchange character instead. Neither inversion involves $K$ or $K'$ directly, and yet the density at the Dirac points also changes afterwards, developing a hexagonal ring of charge around AA that becomes more pronounced as $\alpha$ grows.

\subsection{Berry curvature and Chern numbers}

To define Chern numbers the central bands must be separated from each other at $K$ and $K'$. We add a small staggered potential $V_\delta=(\delta/2)\sigma_z$, with $\delta\approx10$~meV, of the kind produced by an aligned hBN substrate \cite{hunt2013,jung2015}. Since $\sigma_z$ anticommutes with the chiral Hamiltonian, $(H+V_\delta)^2=H^2+\delta^2/4$: the mass shifts all squared energies by the same amount and leaves the eigenstates unchanged, so it opens a gap at the Dirac points without moving the band inversions at $\Gamma$ and $M$. Chern numbers are computed with the Fukui--Hatsugai--Suzuki method on a discretized Brillouin zone \cite{fukui2005}: one defines link variables $U_j(\vec k)=\expval{\Psi(\vec k)|\Psi(\vec k+\hat e_j)}/|\cdots|$, the plaquette field $F_{12}(\vec k)=\ln[U_1U_2U_1^{-1}U_2^{-1}]\in(-\pi,\pi]$, and $C=(2\pi\ii)^{-1}\sum_{\vec k}F_{12}(\vec k)$. This gives an integer for any mesh and is gauge invariant, which matters near an inversion where the states change quickly with $\vec k$. Time reversal makes the total Chern number of both valleys vanish; the values quoted are for one valley.

\begin{figure}[p]
\centering
\includegraphics[width=\linewidth]{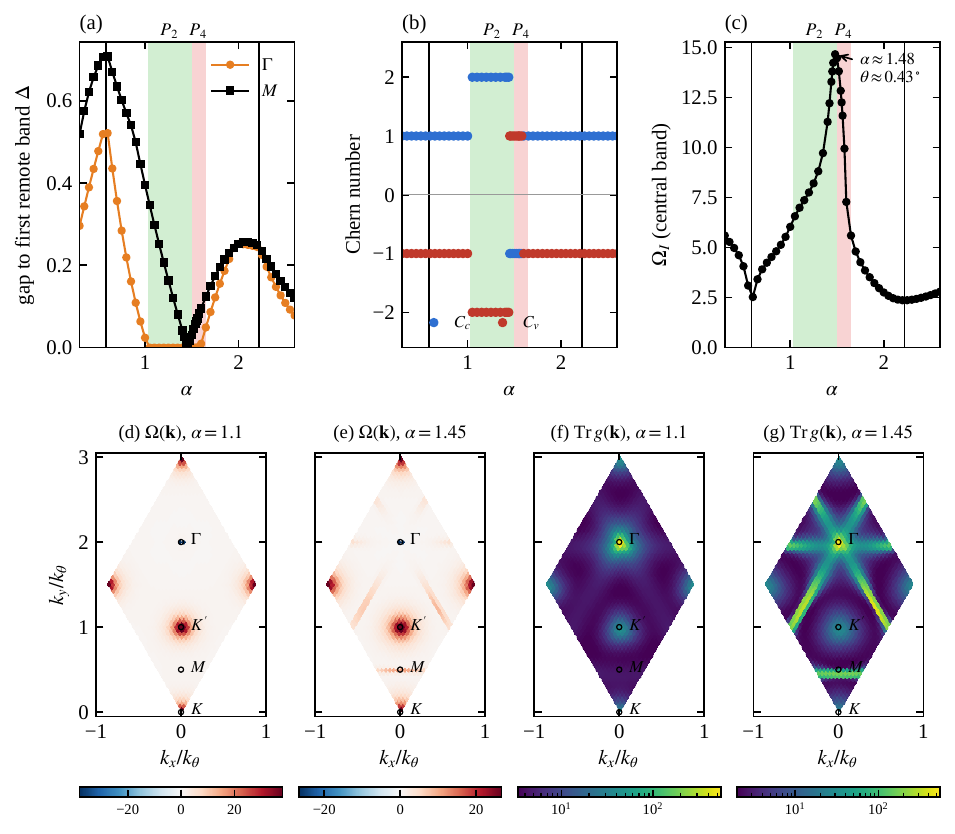}
\caption{Topology of the central bands in the first hybridization window, computed with the plane-wave method and a staggered mass $\delta=10$~meV. (a) Gap between the upper central band and the first remote band at $\Gamma$ and at $M$. Vertical lines: $\alpha_1$ and $\alpha_2$; shaded: phases $P_2$ (green) and $P_4$ (red) of Ref.~\cite{navarro2026a}. (b) Chern numbers of the conduction ($C_c$) and valence ($C_v$) central bands. Outside the window where the upper central band touches the first remote band at $\Gamma$, they are computed with the Fukui--Hatsugai--Suzuki method on a $13\times13$ grid; inside that window we use the values of Ref.~\cite{navarro2026a} (Table~\ref{tab:phases}), obtained by resolving the minigap at $\Gamma$ (see text). (c) Brillouin-zone average $\Omega_I$ of the trace of the quantum metric of the upper central band, with a finer grid ($\Delta\alpha=0.02$) around its maximum at $\alpha\approx1.48$ ($\theta\approx0.43^\circ$). (d),(e) Berry curvature of the upper central band at $\alpha=1.1$ and $1.45$ in the unit cell spanned by $\vec b_1,\vec b_2$ with $K$ at the origin. (f),(g) Trace of the quantum metric at the same couplings (log scale).}
\label{fig:topology}
\end{figure}

At every magic angle the central bands have $\{C_v,C_c\}=\{-1,1\}$ \cite{ledwith2020,ledwith2022}. Between magic angles three other phases appear (Table~\ref{tab:phases}) \cite{navarro2026a}. Phase $P_2$, with $\{-2,2\}$, exists only in the first window, where the bounded overlap channel of Sec.~\ref{sec:qhe} is still comparable to the current channel. Phase $P_3$, with $\{2,-2\}$, appears beyond the second magic angle, and $P_4$, with $\{1,-1\}$, appears in both regimes. Inversions at $M$ produce larger jumps than inversions at $\Gamma$, because the three $M$ points related by $C_3$ invert at the same time and each transfers Berry flux. For example, going from $P_2$ to $P_4$ at $\alpha\approx1.5$ changes $C_c$ from $2$ to $-1$.

Figure~\ref{fig:topology} shows these quantities for the first window. The gap at $\Gamma$ closes at $\alpha\approx1.03$ and stays essentially closed up to $\alpha\approx1.55$, while the gap at $M$ closes and reopens at $\alpha\approx1.46$ (panel a). This is the sequence $\Gamma\to M\to\Gamma$ of Ref.~\cite{navarro2026a}. In the $\Gamma$ window the upper central band hybridizes with the first remote band: at $\Gamma$ they form a degenerate pair, and the Chern number of the central band is defined through the small minigap that separates them \cite{navarro2026a}. With this resolution the central bands carry $C=\pm2$ in $P_2$, jump by $3$ at the $M$ inversion to $C=\mp1$ in $P_4$, and return to $\pm1$ when the $\Gamma$ gap reopens (panel b). One technical remark for anyone repeating the calculation: since a uniform staggered mass satisfies $(H+V_\delta)^2=H^2+\delta^2/4$, it does not by itself separate the two bands at $\Gamma$; a Fukui--Hatsugai--Suzuki calculation that simply follows the lowest branch through the touching point assigns the flux of the quadratic touching to the remote band and returns values shifted by $2$ ($0$ and $-3$). The values of Ref.~\cite{navarro2026a} correspond to assigning that flux to the central band, which follows the character of the states through the band inversion, the central band exchanging its character with the remote band.

The Berry curvature maps show where the topological charge sits. At $\alpha=1.1$ it is concentrated at $K$ and $K'$ (the gapped Dirac points) and at $\Gamma$, where the band touches the remote one. At $\alpha=1.45$, close to the $M$ inversion, additional curvature appears on lines joining $\Gamma$ with the $M$ points, where the central band comes close to the first remote band. This is the precursor of the six points $W_j$ around $\Gamma$ discussed below.

\begin{table}[h]
\centering
\caption{Topological phases of the central bands in the first three hybridization windows \cite{navarro2026a}.}
\label{tab:phases}
\begin{tabular}{@{}lcc@{}}
\toprule
Phase & $\alpha$-interval & $[C_v,C_c]$\\ \midrule
$P_2$ & $[1.030,\,1.500]$ & $[-2,\,2]$ \\
$P_4$ & $[1.500,\,1.650]$ & $[1,\,-1]$ \\
$P_3$ & $[3.180,\,3.250]$ & $[2,\,-2]$ \\
$P_4$ & $[3.250,\,3.310]$ & $[1,\,-1]$ \\
$P_3$ & $[4.730,\,4.913]$ & $[2,\,-2]$ \\
$P_4$ & $[4.913,\,4.915]$ & $[1,\,-1]$ \\
\bottomrule
\end{tabular}
\end{table}

The local Berry flux $C_n(\vec k)=(2\pi)^{-1}\vec S_g(\vec k)\cdot\vec\Omega_n(\vec k)$ shows where the topological charge goes. At $K$ and $K'$ it is $\pm1/2$ for any $\alpha$. At an $M$-point inversion the six points $W_j$ around $\Gamma$ carry $C_0(W_j)=-1/2$ (central band), $C_1(W_j)=+1/2$ (first remote band) and $C_2(W_j)\approx0$, so that $\sum_n\sum_jC_n(W_j)\approx0$. At a $\Gamma$-point inversion, three points each contribute $1/3$, giving $C_0(\Gamma)\approx1$, $C_1(\Gamma)\approx-1$ and $C_2(\Gamma)\approx0$. These sum rules are the analogue of the Chern-number bookkeeping at band touchings in the Hofstadter butterfly \cite{naumis2016,naumis2019}. The touching points are also Van Hove singularities, where the density of states is enhanced.

\subsection{Quantum metric and superfluid weight}

The quantum (Fubini--Study) metric $g^n_{\alpha\beta}(\vec k)$ (Appendix~\ref{app:quantummetric}) complements the Berry curvature and has been useful in other flat-band models \cite{espinosa2023}. Its Brillouin-zone average,
\begin{equation}
\Omega_I = \frac{1}{N_c}\sum_n\sum_{\vec k} \Tr[g^n(\vec k)], \qquad \Tr[g^n]=g^n_{xx}+g^n_{yy},
\label{eq:MVcumulant}
\end{equation}
is the gauge-invariant part of the Marzari--Vanderbilt spread and gives a lower bound on the spread of the Wannier functions \cite{marzari1997}. In TBG it is smallest at the magic angles, where the whole flat band is generated by one analytic function through Eq.~\eqref{eq:flatbandfactor}, and it peaks sharply at the band touchings, more strongly at $M$ (linear) than at $\Gamma$ (quadratic) \cite{navarro2026a}. Figure~\ref{fig:topology}(c) shows this: $\Omega_I$ has sharp minima at $\alpha_1$ and $\alpha_2$, grows through the $P_2$ window, and peaks at the $M$ inversion, in agreement with Ref.~\cite{navarro2026a}. The maps in Fig.~\ref{fig:topology}(f,g) show that the metric is large near $\Gamma$ and, at $\alpha=1.45$, along the same lines where the Berry curvature appears.

The maximum of $\Omega_I$ deserves attention. A finer scan places it at $\alpha\approx1.48$, i.e., $\theta\approx0.43^\circ$, just after the gap at $M$ closes ($\alpha\approx1.46$) and at the boundary between $P_2$ and $P_4$. There $\Omega_I\approx14.7$, about six times its value at the first magic angle ($\Omega_I\approx2.5$). Several ingredients meet at this point. (i) The central band touches the first remote band at the three $M$ points, which are saddle points of the dispersion, so the density of states has Van Hove singularities close to the band edge. (ii) The central band is still narrow compared with the Coulomb scale, since the whole low-energy manifold remains squeezed for $\alpha\gtrsim1$. (iii) The band is topological, $|C|=2$ on the $P_2$ side, and its quantum geometry is maximal. Through Eq.~\eqref{eq:Dsconvgeom}, a large quantum metric raises the geometric part of the superfluid weight, and the minimal-metric bound for a band with Chern number $C$, $\int\Tr g\,\dd^2k\ge2\pi|C|$, is larger for $|C|=2$ than for the usual $|C|=1$ \cite{peotta2015,huhtinen2022}. The combination of a high density of states, strong quantum geometry and higher Chern number makes the region $1.4\lesssim\alpha\lesssim1.55$ ($\theta\approx0.41^\circ$--$0.45^\circ$) a natural place to look for correlated phases beyond those of the first magic angle: geometric superconductivity, Chern insulators with $|C|=2$ at integer fillings, and possibly fractional Chern states if the band can be made flatter, e.g., with a perpendicular displacement field or pressure. Two cautions apply. The band is not flat there, so the kinetic energy competes with the interaction and the quantum-metric contribution to pairing is not the only one; and at such small angles lattice relaxation becomes important, so the single-particle window found here should be checked beyond the chiral limit. With these caveats, this is the configuration we would study next.

This connects to superconductivity. In a multiband superconductor the mean-field superfluid weight has a conventional part and a geometric part \cite{peotta2015,huhtinen2022},
\begin{equation}
\begin{gathered}
D_s^{\mu\nu} = D_s^{\mu\nu,\rm conv} + D_s^{\mu\nu,\rm geom}, \\
D_s^{\mu\nu,\rm conv}\propto\sum_n\int \partial_\mu\partial_\nu E_n(\vec k)\,f_n(\vec k)\,\dd^2k, \qquad
D_s^{\mu\nu,\rm geom}\propto\Delta\sum_n\int g_n^{\mu\nu}(\vec k)\,\dd^2k,
\end{gathered}
\label{eq:Dsconvgeom}
\end{equation}
where $\Delta$ is the pairing gap and $f_n$ the occupation. For an exactly flat band the first term vanishes and only the geometric one survives. Phase $P_2$ combines $|C|=2$ with a maximum of the quantum metric, which makes it an interesting candidate for geometric superconductivity and, with strong interactions, for fractional Chern insulators \cite{xie2021,ledwith2022}. Note also that $\Omega_I$ is smallest precisely where the bandwidth vanishes, so the two ingredients of flat-band superconductivity, small kinetic energy and large quantum geometry, are optimized at slightly different angles.

\section{Conclusions and Outlook}
\label{sec:conclusion}

The common thread of these notes is the squared chiral Hamiltonian,
\begin{equation}
H^2 = -\nabla^2 + \alpha^2\hat A^2 - 2\ii\alpha(\hat A\cdot\nabla) + \alpha\hat F_{xy},
\tag{\ref{eq:H2enlightening} revisited}
\end{equation}
an electron in a non-Abelian $SU(2)$ pseudo-magnetic field produced by the interlayer tunneling. Looking at it in different ways, we have discussed the following points.

\begin{itemize}[leftmargin=1.4em]
\item The energy sum rule separates kinetic, confinement and interlayer energies. The interlayer term has a bounded overlap part and an unbounded current part; their ratio grows as $\alpha^2$, and they are comparable only for $\alpha\lesssim1$. This is why the first magic angle differs from the others.
\item The $3/2$ rule follows from rescaling the zero-mode equation, which turns $\alpha$ into a stretching of the moir\'e potential, together with the fact that the potential is periodic only on a three-cell supercell. Each parity family repeats with period $3$ in $\alpha$, and the two families are interleaved.
\item At the magic angles kinetic and confinement energies are equal, and the zero modes carry an orbital angular momentum that grows by one unit per magic angle. For large $\alpha$ they become coherent Landau states of width $\sigma=1/\sqrt{3\alpha}$: near the AA point the zero-mode problem is a lowest-Landau-level problem in an effective field $B_{\rm eff}=3\alpha$, with the electrons sitting at displaced guiding centres.
\item Deforming the coupling into an Abelian one removes the magic-angle sequence, which shows that the non-Abelian structure is essential.
\item Between magic angles, band inversions at $\Gamma$ and $M$ produce Chern numbers $C=\pm2$ in narrow windows, where the quantum metric is also maximal.
\end{itemize}

Several questions remain open. The squared-Hamiltonian approach should apply to other moir\'e systems with a chiral limit, such as twisted multilayer graphene, where flat bands with higher Chern numbers are known \cite{ledwith2022}. It would be interesting to see whether a similar rescaling argument fixes their magic-angle sequences, and whether their gauge structure goes beyond $SU(2)$. A second direction is to use the coherent-state basis of Sec.~\ref{sec:localization}, labelled by the angular momentum $m$, as a starting point for projecting the Coulomb interaction, and to compare the resulting pairing estimates with the geometric superfluid weight of Eq.~\eqref{eq:Dsconvgeom}. Third, real samples are not chiral: the AA tunneling is reduced by relaxation but not removed. How the structures described here change for finite $w_{AA}$, and why the first magic angle is the one that survives, is the subject of a separate work \cite{navarro2026b}. Finally, the $|C|=2$ windows between the first and second magic angles are, in principle, accessible to local compressibility, scanning tunneling and magnetotransport measurements at small twist angles. The consequences of the flat-band structure for quantum information, including entanglement, quantum Fisher information, thermal robustness and control by an external magnetic field, are developed in a companion work \cite{qipaper}. More generally, the strategy of using a chiral symmetry to square a Hamiltonian and read off the resulting gauge field may be useful in other square-root topological problems \cite{navarroqip2025}.

\appendix

\section{Plane-Wave Method}
\label{app:bloch}

We expand $\Psi_{\vec k}(\vec r)=\sum_{mn}(a_{mn},b_{mn}e^{\ii\vec q_1\cdot\vec r})^Te^{\ii(\vec K_{mn}+\vec k)\cdot\vec r}$, with $\vec K_{mn}=m\vec b_1+n\vec b_2$, and similarly for $\chi$ with coefficients $c_{mn},d_{mn}$. Substituting into $H\Phi=E\Phi$ gives \cite{navarro2023a}
\begin{align}
(K'^x_{mn}+\ii K'^y_{mn})a_{mn} + \alpha\big(b_{mn}+e^{\ii\phi}b_{m+1,n}+e^{-\ii\phi}b_{m,n+1}\big) &= E\,a_{mn}, \label{eq:recA}\\
\big(K'^x_{mn}+\ii(K'^y_{mn}+q_1)\big)b_{mn} + \alpha\big(a_{mn}+e^{\ii\phi}a_{m-1,n}+e^{-\ii\phi}a_{m,n-1}\big) &= E\,b_{mn}, \label{eq:recB}
\end{align}
with $\vec K'_{mn}=\vec K_{mn}+\vec k$, and an analogous pair for $c_{mn},d_{mn}$. Keeping $|m|,|n|\le N$ gives a matrix of dimension $4(2N+1)^2$. Because $H$ is chiral, its eigenvalues are $\pm$ the singular values of the matrix of $D$, which halves the problem; the same matrix gives $H^2=D^\dagger D$. The figures in these notes were computed this way, with a circular cutoff $|\vec K'_{mn}|<P$ and $P\approx2.2\alpha+7$, which is enough to converge the magic angles of Table~\ref{tab:magicangles} to three decimals. The magic angles themselves were obtained from the eigenvalues of the operator $(-\ii\bar\partial)^{-1}\mathcal U$, where $\mathcal U$ is the off-diagonal part of $D$: a zero mode at coupling $\alpha$ corresponds to the eigenvalue $-1/\alpha$ \cite{tarnopolsky2019}. For the interpolated coupling of Sec.~\ref{sec:abelian}, the term $U(-\vec r)$ connects $\vec k$ with $\vec k+\vec q_\nu$, which does not belong to the same coset of the reciprocal lattice, so the basis must include the whole lattice generated by the $\vec q_\nu$.

At $\Gamma$ the problem can be reduced further using $C_3$. The reciprocal vectors transform as, 
\begin{equation}
    R_\phi(\vec b_1)=\vec b_2-\vec b_1, R_\phi(\vec b_2)=-\vec b_1, R_{2\phi}(\vec b_1)=-\vec b_2, R_{2\phi}(\vec b_2)=\vec b_1-\vec b_2.
 \end{equation}
Combined with $\Psi_{\vec k}(R_\phi\vec r)=e^{-\ii\phi}\Psi_{\vec k}(\vec r)$ at $\vec k=\vec q_1$ this gives,
\begin{equation}
e^{-\ii\phi}a_{mn} = a_{n,-m-n+1}, \qquad e^{\ii\phi}a_{mn} = a_{-m-n+1,m},
\label{eq:C3coeffrelations}
\end{equation}
and, together with $b_{mn}=\ii\mu_\alpha a_{1-m,1-n}$ from Eq.~\eqref{eq:parityrelation}, the reduced recursion at the magic angles \cite{navarro2023a}
\begin{equation}
\left(\ii\frac{\sqrt3}{2}(m-n)+1-\frac32(m+n)\right)a_{mn} \mp \alpha_m\big(a_{1-m,1-n}+a_{-n,m+n}+a_{m+n,-m}\big) = 0.
\label{eq:reducedrecursion}
\end{equation}
The combination in parentheses is a Fourier coefficient plus its two $C_3$ images.

\subsection*{Solving the recursion for small $\alpha$}

Equation~\eqref{eq:reducedrecursion} can be solved by hand for small $\alpha$, and doing so shows how the zero mode at $\Gamma$ is built shell by shell. Write the diagonal coefficient as
\begin{equation}
c_{mn}\equiv \ii\frac{\sqrt3}{2}(m-n)+1-\frac32(m+n) = \ii P_{mn}, \qquad P_{mn}=\big(\vec K_{mn}+\vec q_1\big)_x+\ii\big(\vec K_{mn}+\vec q_1\big)_y,
\label{eq:cmn}
\end{equation}
so that $|c_{mn}|$ is the distance of the plane wave $e^{\ii(\vec K_{mn}+\vec q_1)\cdot\vec r}$ from the origin of momentum space, i.e., its kinetic energy. Taking the sign that corresponds to $\mu_\alpha=+1$ (the parity of $\alpha_1$, Table~\ref{tab:magicangles}), Eq.~\eqref{eq:reducedrecursion} reads
\begin{equation}
a_{mn} = \frac{\alpha}{c_{mn}}\,S_{mn}, \qquad S_{mn}\equiv a_{1-m,1-n}+a_{-n,m+n}+a_{m+n,-m}.
\label{eq:recursionsolved}
\end{equation}
Each coefficient is fed by the coefficient at the inverted site $(1-m,1-n)$ and by its two $C_3$ images, weighted by $\alpha/c_{mn}$. Since $|c_{mn}|$ grows with the distance from the origin, for small $\alpha$ the coefficients decay quickly with $|\vec K_{mn}+\vec q_1|$, and the wavefunction can be built by organizing the plane waves into shells of increasing $|c_{mn}|$ (Fig.~\ref{fig:shells}).

\begin{figure}[t]
\centering
\begin{tikzpicture}[scale=1.25,>=stealth]
  \def\sq{0.8660254}
  \foreach \m in {-2,...,3}{
    \foreach \n in {-2,...,3}{
      \pgfmathsetmacro{\x}{\sq*(\m-\n)}
      \pgfmathsetmacro{\y}{1.5*(\m+\n)-1}
      \pgfmathsetmacro{\rr}{sqrt(\x*\x+\y*\y)}
      \ifdim \rr pt < 3.1pt
        \fill[gray!55] (\x,\y) circle (1.6pt);
      \fi
    }
  }
  \draw[gray!60,dashed] (0,0) circle (1);
  \draw[gray!60,dashed] (0,0) circle (2);
  \draw[gray!60,dashed] (0,0) circle (2.6458);
  \foreach \x/\y/\l/\a in {0/-1/{(0,0)}/below, 0.8660254/0.5/{(1,0)}/right, -0.8660254/0.5/{(0,1)}/left}{
    \fill[red!80!black] (\x,\y) circle (3pt);
    \node[\a,red!70!black,font=\footnotesize] at (\x,\y) {$\l$};
  }
  \foreach \x/\y/\l/\a in {0/2/{(1,1)}/above, 1.7320508/-1/{(1,-1)}/below right, -1.7320508/-1/{(-1,1)}/below left}{
    \fill[blue!70!black] (\x,\y) circle (3pt);
    \node[\a,blue!70!black,font=\footnotesize] at (\x,\y) {$\l$};
  }
  \draw[black] (0,0.5) node{$\times$};
  \node[right,font=\scriptsize] at (0.05,0.62) {$(\tfrac12,\tfrac12)$};
  \draw[<->,thick,blue!70!black,shorten >=4pt,shorten <=4pt] (0,2) -- (0,-1);
  \draw[->,thick,red!80!black] (0.12,-1.1) arc (-60:240:0.24);
  \node[font=\footnotesize] at (0,-3.35) {$|c_{mn}|=1$ (red, order $\alpha^0$), \ $2$ (blue, order $\alpha$), \ $\sqrt7$ (grey, order $\alpha^2$)};
  \draw[->,gray] (-2.9,0)--(2.9,0) node[right,font=\footnotesize]{$p_x$};
  \draw[->,gray] (0,-2.9)--(0,2.9) node[above,font=\footnotesize]{$p_y$};
\end{tikzpicture}
\caption{Plane waves $e^{\ii(\vec K_{mn}+\vec q_1)\cdot\vec r}$ entering the zero mode at $\Gamma$, plotted at their momenta (units of $k_\theta$) and labelled by $(m,n)$. Dashed circles mark $|c_{mn}|=1,2,\sqrt7$. The recursion~\eqref{eq:recursionsolved} couples each site to its image under inversion through $(\tfrac12,\tfrac12)$ (cross) and to the two $C_3$ images of that site. The zero-order shell (red) is mapped by inversion onto the first-order shell (blue), and site $(0,0)$ is coupled to itself by the two $C_3$ terms (red loop).}
\label{fig:shells}
\end{figure}

\emph{Order $\alpha^0$.} The three plane waves with the smallest kinetic energy, $|c_{mn}|=1$, are $(m,n)=(0,0),(1,0),(0,1)$, with momenta $\vec q_1,\vec q_2,\vec q_3$. They form one $C_3$ orbit, and Eq.~\eqref{eq:C3coeffrelations} fixes their relative phases, $a_{10}=e^{\ii\phi}a_{00}$ and $a_{01}=e^{2\ii\phi}a_{00}$. Hence
\begin{equation}
\psi_{\Gamma,1}^{(0)}(\vec r) = a_{00}\big(e^{\ii\vec q_1\cdot\vec r}+e^{\ii\phi}e^{\ii\vec q_2\cdot\vec r}+e^{-\ii\phi}e^{\ii\vec q_3\cdot\vec r}\big) = a_{00}\,U(-\vec r),
\label{eq:order0}
\end{equation}
the zero-order state quoted in Appendix~\ref{app:pert} and in Sec.~\ref{sec:rescaling}.

\emph{Order $\alpha$.} Inversion through $(\tfrac12,\tfrac12)$ maps the zero-order shell onto the sites $(1,1),(0,1),(1,0)$, so the only new sites reached at first order are the $C_3$ orbit of $(1,1)$, namely $(1,1),(1,-1),(-1,1)$, with $c_{11}=-2$ and momenta $-2\vec q_1,-2\vec q_3,-2\vec q_2$. Equation~\eqref{eq:recursionsolved} at $(1,1)$ gives
\begin{equation}
a_{11} = \frac{\alpha}{c_{11}}\big(a_{00}+a_{-1,2}+a_{2,-1}\big) \simeq -\frac{\alpha}{2}\,a_{00},
\label{eq:a11}
\end{equation}
because $a_{-1,2}$ and $a_{2,-1}$ belong to the $|c_{mn}|=\sqrt7$ shell and are $\mathcal O(\alpha^2)$. The $C_3$ relations give $a_{1,-1}=e^{-\ii\phi}a_{11}$, $a_{-1,1}=e^{-2\ii\phi}a_{11}$, and the three plane waves add up to $U(2\vec r)$. To first order,
\begin{equation}
\psi_{\Gamma,1}(\vec r) \simeq a_{00}\Big[U(-\vec r)-\frac{\alpha}{2}\,U(2\vec r)\Big]+\mathcal O(\alpha^2),
\qquad
\psi_{\Gamma,2}(\vec r)=\ii\,\psi_{\Gamma,1}(-\vec r).
\label{eq:order1}
\end{equation}
This is the self-consistent zero mode. It should not be confused with the expansion of the lowest (nonzero-energy) level at $\Gamma$ for generic $\alpha$ given in Appendix~\ref{app:pert}, which carries a different energy denominator.

\emph{Closing the recursion: estimate of $\alpha_1$.} The zero-order shell must also satisfy Eq.~\eqref{eq:recursionsolved}. At $(0,0)$, with $c_{00}=1$, the inverted site is $(1,1)$ and both $C_3$ images are $(0,0)$ itself (red loop in Fig.~\ref{fig:shells}):
\begin{equation}
a_{00} = \alpha\big(a_{11}+2a_{00}\big).
\label{eq:closure}
\end{equation}
The equations at $(1,0)$ and $(0,1)$ are the same by $C_3$. Keeping only the zero-order shell ($a_{11}=0$) gives the crude estimate $\alpha_1\approx1/2$. Inserting the first-order result~\eqref{eq:a11},
\begin{equation}
1 = 2\alpha-\frac{\alpha^2}{2}
\quad\Longrightarrow\quad
\alpha_1\simeq 2-\sqrt2 = 0.5858,
\label{eq:alpha1estimate}
\end{equation}
which agrees with the exact value $\alpha_1=0.5857$ to four digits. The other sign, $\mu_\alpha=-1$, gives $\alpha^2/2+2\alpha+1=0$, with no positive root, so the first magic angle must have even parity, as in Table~\ref{tab:magicangles}. As a check, the numerical zero mode at $\alpha_1$ (plane-wave basis with $|m|,|n|\le12$) gives $a_{11}/a_{00}=-0.2925$, against $-\alpha_1/2=-0.2928$ from Eq.~\eqref{eq:a11}, and $|a_{-1,2}/a_{00}|=0.064$, consistent with an $\mathcal O(\alpha^2)$ coefficient.

The same procedure can be continued: at order $\alpha^2$ the $\sqrt7$ shell is generated from $(1,1)$ and feeds back into $(1,1)$, and so on. The convergence of this expansion is controlled by $\alpha/|c_{mn}|$; for $\alpha\gg1$ many shells contribute with comparable weight, and the asymptotic analysis of Appendix~\ref{app:pert} and Sec.~\ref{sec:rescaling} takes over.

\section{Perturbative and Asymptotic Expansions}
\label{app:pert}

\subsection{Zero modes at small $\alpha$}

At $\Gamma$ the zero mode has the expansion $\psi_{\Gamma,1}(\vec r)=U(-\vec r)+\tfrac{\alpha}{3}U(2\vec r)+\mathcal O(\alpha^2)$, $\psi_{\Gamma,2}=\ii\mu_\alpha\psi_{\Gamma,1}(-\vec r)$. At zeroth order $-\nabla^2U(-\vec r)=U(-\vec r)$, since $|\vec q_\nu|=1$, so $E_0^2=1$. At first order one needs the action of the two parts of the interlayer operator, $A_g^\dagger(\vec r)U(\vec r)=3\ii U(-\vec r)$ and $A_f^\dagger(\vec r)U(\vec r)=-\ii(U(2\vec r)-U(-\vec r))$, which follow from $\hat q_l^\perp\cdot\hat q_s=\pm\sqrt3/2$ for $l\ne s$. This gives $\expval{T}=1$, $\expval{V}=0$, $\expval{A_f}=-\alpha$ and $\expval{A_g}=-3\alpha$, the values used in Eq.~\eqref{eq:EGammaapprox}.

At $K$ the expansion $\Psi_K=(1+\alpha^2u_2+\cdots,\ \alpha u_1+\alpha^3u_3+\cdots)^T$, with $u_1=-\ii\sum_\mu e^{\ii\vec q_\mu\cdot\vec r}$ and normalization $N=\tfrac{8\pi^2}{3\sqrt3}(1+3\alpha^2+2\alpha^4+\tfrac67\alpha^6+\cdots)$, gives \cite{navarro2022}
\begin{equation}
\expval{T}\approx\expval{V}\approx3\alpha^2-3\alpha^4, \qquad \expval{A_f}\approx-6\alpha^2+12\alpha^4, \qquad \expval{A_g}\approx-6\alpha^4+18\alpha^6.
\label{eq:appKvalues}
\end{equation}
This confirms $\expval{T}=\expval{V}$ and $\expval{T}+\expval{V}+\expval{\mathcal A}\approx0$ to order $\alpha^4$, and gives $\expval{A_g}/\expval{A_f}\approx\alpha^2$.

\subsection{Coherent states for $\alpha\to\infty$}

In complex notation the coupling is $U(z,z^*)=\sum_\mu(\ii Q_\mu)e^{-\ii(Q_\mu^*z+Q_\mu z^*)/2}$, with $Q_\mu=q_{\mu,x}+\ii q_{\mu,y}$. The zero-mode equation at a magic angle is $2\partial_{z^*}\psi_\mp=\pm\alpha_mU\psi_\mp(-z,-z^*)$. Expanding the exponentials to first order and using $\sum_\mu Q_\mu=0$ and $\sum_\mu|Q_\mu|^2=3$ gives \cite{navarro2023a}
\begin{equation}
2\partial_{z^*}\psi_\mp(z,z^*) \approx \Big(\frac{3\alpha_m}{2}\Big)(-z)\,\psi_\mp(z,z^*).
\label{eq:coherentderivation}
\end{equation}
The antiholomorphic derivative of $f_\lambda(z)e^{-|z|^2/4\sigma^2}$ is $-(z/2\sigma^2)$ times the same function, with $f_\lambda(z)$ analytic. Comparing with Eq.~\eqref{eq:coherentderivation} fixes $\sigma^2=1/(3\alpha_m)$. Summing over the three centres $Q_\mu$ and their images $-Q_\mu$ with the sign fixed by the parity,
\begin{equation}
\psi_\pm(z,z^*) = \frac{1}{\sqrt6}\sqrt{\frac{3\alpha}{2\pi}}\sum_{\mu=1}^3\Big(f_{Q_\mu}(z) \pm f_{-Q_\mu}(z)\Big)e^{-|z|^2/4\sigma^2}, \qquad \sigma=\frac{1}{\sqrt{3\alpha}},
\label{eq:appcoherentfinal}
\end{equation}
with $f_\lambda(z)=\exp[\tfrac{1}{2\sigma^2}\lambda^*z-\tfrac{1}{4\sigma^2}|\lambda|^2]$. The overlap between Gaussians centred at different points is neglected, which is justified for $\sigma\to0$. No parameter is fitted to the numerics.

\section{Equipartition and the Off-Diagonal Sum Rule}
\label{app:equipartition}

Apply $\partial$ to the zero-mode equation $\bar\partial\psi_1(\vec r)=\alpha_m\mu_\alpha U(\vec r)\psi_1(-\vec r)$:
\begin{equation}
\partial\bar\partial\psi_1(\vec r) = \alpha_m\mu_\alpha\big[\partial U(\vec r)\,\psi_1(-\vec r) + U(\vec r)\,\partial\psi_1(-\vec r)\big].
\label{eq:appderivativestep}
\end{equation}
Using that $\partial\bar\partial=\nabla^2$, the equation for the other sublattice sector, $D^\dagger(-\vec r)\chi=0$, and the $\Gamma$-point relations $\chi_1(\vec r)=\mu_\alpha\psi_2(\vec r)=\ii\mu_\alpha^2\psi_1(-\vec r)$, the right-hand side can be written in terms of $\psi_\pm$ alone, resulting in \cite{navarro2023a},
\begin{equation}
-\nabla^2\psi_\pm(\vec r) = \big(\alpha_m^2|U(\vec r)|^2 + \alpha_m\,\partial U(\vec r)\big)\psi_\pm(\vec r).
\label{eq:appsecondordereq}
\end{equation}
The expectation value of $\partial U$ vanishes by symmetry, and we obtain Eq.~\eqref{eq:equipartition}. For $m\to\infty$ the state is concentrated at six centres (three per layer); normalizing over them gives $\expval{|U|^2}_m\to\tfrac{1}{12}\sum_\mu\big[|U(\vec q_\mu)|^2+|U(-\vec q_\mu)|^2\big]\approx1$, so that $\expval{T}_m\approx\alpha_m^2$.

The off-diagonal sum rule follows from $H^2\Psi=0$. The diagonal part of $\expval{H^2}$ is $\expval{T}+\expval{V}$, and it must be compensated by the off-diagonal part, $\alpha\big(\expval{1|A^\dagger|2}+\expval{2|A|1}\big)=-(\expval{T}+\expval{V})$. With equipartition and the normalization above this gives Eq.~\eqref{eq:offdiagquant}, $\expval{1|A^\dagger|2}_m+\expval{2|A|1}_m=-\alpha_m$ \cite{navarro2023a}.

\section{The Harmonic-Oscillator Limit and the Origin}
\label{app:HOlimit}

Here we collect the steps behind Sec.~\ref{sec:LLL} and discuss where the oscillator description holds.

\emph{Derivation.} In complex notation $\partial_z=\tfrac12(\partial_x-\ii\partial_y)$, $p_z=-\ii\partial_z$, and $4p_zp_z^*=p_x^2+p_y^2$. The diagonal block of $H^2$ contains $\alpha^2|U(-\vec r)|^2$, which with Eq.~\eqref{eq:Ulinear} becomes $(3\alpha/2)^2|z|^2$. In the off-diagonal block, $A^\dagger=A_g^\dagger+A_f^\dagger$; expanding the exponentials to first order and using $Q^\perp_\mu=\ii Q_\mu$ gives $A_g^\dagger\psi(-z)\approx3(zp_z-z^*p_{z^*})\psi(-z)=-3\ii L_z\psi(-z)$ and $A_f^\dagger\psi(-z)\approx-3\ii\psi(-z)$ \cite{navarro2023a}. With $\psi_2(\vec r)=\ii\mu_\alpha\psi_1(-\vec r)$ and a state of definite parity this yields Eq.~\eqref{eq:Landau}. Writing $\vec\Pi=\vec p+\vec{\mathcal A}$ with $\vec{\mathcal A}=\tfrac{B}{2}(-y,x)$, one has $\Pi^2=p^2+\tfrac{B^2}{4}r^2-BL_z$, so Eq.~\eqref{eq:Landau} is $\Pi^2\psi=B\psi$ with $B=3\alpha$. Since $[\Pi_x,\Pi_y]=-\ii B$, the operator $a=(\Pi_x-\ii\Pi_y)/\sqrt{2B}$ satisfies $[a,a^\dagger]=1$ and $\Pi^2=B(2a^\dagger a+1)$. The guiding-centre coordinates commute with $\vec\Pi$ and generate the degeneracy of each level. The zero modes are the states with $a\psi=0$, Eq.~\eqref{eq:LLLform}.

\emph{Why the expansion is made at $\vec r=0$.} The AA point is the only point where $U$ vanishes in the moir\'e cell, and it vanishes linearly, Eq.~\eqref{eq:Ulinear}. A linear zero of the coupling is exactly what produces a uniform effective field, in the same way as a linearly growing vector potential does in the symmetric gauge. The origin therefore fixes the gauge and the value $B_{\rm eff}=3\alpha$, not the location of the electrons.

\emph{The problem at $\vec r=0$.} If Eq.~\eqref{eq:Landau} is read as an oscillator centred at the origin, its ground state would be a Gaussian at AA, in contradiction with Fig.~\ref{fig:density}, where the density at AA is depleted and the charge sits on a ring of radius $\approx\pi/3$. The contradiction disappears once one remembers that the oscillator is a Landau problem: the lowest level is degenerate, and a guiding-centre displacement $\lambda$ costs no energy. The analytic prefactor $f(z)$ selects the guiding centres, and in TBG they are fixed by the moir\'e potential beyond linear order and by the Bloch boundary conditions at $\lambda\approx\pm(\pi/3)\vec q_\mu$. Around each centre, shifting $z\to z-\lambda$ turns Eq.~\eqref{eq:Landau} into the same problem centred at $\lambda$, so the oscillator is effectively centred at the localization centres. In addition, the antisymmetric combination, which is the zero mode for one of the two parities, vanishes identically at $\vec r=0$.

\emph{Range of validity.} The linear form of $U$ is exact only for $|\vec r|\ll1$. Along the directions of $\pm\vec q_\mu$ the error is moderate at $|\vec r|=\pi/3$ ($|U|^2=2$ instead of $2.25$, Fig.~\ref{fig:LLL}a), but farther from AA the exact $|U|^2$ is bounded (it never exceeds $9$) and oscillates, so the parabola overestimates the confinement there (Fig.~\ref{fig:LLL}a). What the zero mode actually probes is a neighbourhood of size $\ell_B=1/\sqrt{3\alpha}$ around each centre, which shrinks with $\alpha$ ($\ell_B=0.16$ at $\alpha_9$). The oscillator picture is therefore an asymptotic statement for $\alpha\gg1$. Its accuracy is measured by Fig.~\ref{fig:coherent}(b): $5$--$15\%$ deviations in the width for $\alpha_2$ to $\alpha_4$, and below $5\%$ from $\alpha_5$ on. It does not apply to the first magic angle, where $\ell_B\approx0.75$, the Gaussians of different centres overlap strongly, and the interlayer overlap channel $A_f$ is as important as the current channel $A_g$ (Sec.~\ref{sec:qhe}). Finally, the oscillator has lowest-Landau-level solutions for any $\alpha$, so it cannot produce the magic-angle sequence by itself; that requires the global boundary condition of Sec.~\ref{sec:rescaling}.

\section{Polar Form of the Interlayer Operator}
\label{app:polar}

In polar coordinates $(r,\nu)$ around a localization centre,
\begin{equation}
A^\dagger(\vec r) = \sum_{\mu=1}^3 e^{-\ii\vec q_\mu\cdot\vec r}\Big[2\big(\hat q_\mu^\perp\cdot\hat e_r\,P_r + \hat q_\mu^\perp\cdot\hat e_\nu\,\tfrac{L_\nu}{r}\big) - \ii\Big],
\label{eq:appAdaggerpolar}
\end{equation}
with $P_r=-\ii\partial_r$ and $L_\nu=-\ii\partial_\nu$. At the centres $\pm(\pi/3)\vec q_\mu$, where $r\approx\pi/3$, $\hat q_\mu^\perp\cdot\hat e_r=0$ and $\hat q_\mu^\perp\cdot\hat e_\nu=1$, so
\begin{equation}
A^\dagger(\vec r) \approx A_0^\dagger(\vec r) + \sum_{\mu=1}^3 e^{-\ii\vec q_\mu\cdot\vec r}(2L_\nu-\ii),
\label{eq:appAdaggerapprox}
\end{equation}
where $A_0^\dagger$ contains the radial momentum. Table~\ref{tab:A0convergence} shows that its matrix element is small and decreases with $m$ \cite{navarro2023a}. Dropping it and using Eq.~\eqref{eq:offdiagquant} leads to Eq.~\eqref{eq:Lnuquant}.

\begin{table}[h]
\centering
\caption{Matrix element of the radial part of the interlayer operator at the magic angles \cite{navarro2023a}.}
\label{tab:A0convergence}
\begin{tabular}{@{}cccccc@{}}
\toprule
$m$ & 5 & 6 & 7 & 8 & 9\\ \midrule
$\expval{1|A_0^\dagger|2}_m$ & 0.419 & 0.257 & 0.211 & 0.179 & 0.164\\
\bottomrule
\end{tabular}
\end{table}

\section{Quantum Metric}
\label{app:quantummetric}

The distance between neighbouring Bloch states, $D^2=1-|\expval{\Psi(\vec k)|\Psi(\vec k+\dd\vec k)}|^2$, expanded to second order in $\dd\vec k$, defines the Fubini--Study metric, $D^2=\sum_{\alpha\beta}g_{\alpha\beta}(\vec k)\dd k_\alpha\dd k_\beta$, with
\begin{equation}
g_{\alpha\beta}(\vec k) = \Re\Big(\expval{\partial_\alpha\Psi|\partial_\beta\Psi}-\expval{\partial_\alpha\Psi|\Psi}\expval{\Psi|\partial_\beta\Psi}\Big)
= \Re\expval{\partial_\alpha\Psi|Q_{\vec k}|\partial_\beta\Psi},
\label{eq:gabprojector}
\end{equation}
where $Q_{\vec k}=1-\ket{\Psi(\vec k)}\bra{\Psi(\vec k)}$. The trace of the metric measures how fast the state changes with $\vec k$. Its Brillouin-zone integral is the gauge-invariant part $\Omega_I$ of the Marzari--Vanderbilt spread of the Wannier functions \cite{marzari1997}. At a magic angle all flat-band states are generated from $\psi_K$ by Eq.~\eqref{eq:flatbandfactor}, so the metric is small; at a band touching the character of the state changes abruptly, and the metric peaks.

\section{A Mathieu-Equation Analogy}
\label{app:mathieu}

The migration of the Fourier weight to higher harmonics can be pictured with a parametric oscillator, $\ddot\psi+(a-2q\cos2t)\psi=0$, with $a=3\alpha^2$ and $q=\alpha^2/2$ chosen to mimic $\alpha^2A^2(\vec r)$ along one direction. Its Floquet solutions are unstable inside tongue-shaped regions that start at $a=n^2$ for $q=0$,
\begin{equation}
a_\pm(q) \approx n^2 \pm \frac{q^n}{2^{n-1}[(n-1)!]^2} + \mathcal O(q^{n+1}).
\label{eq:tongueboundary}
\end{equation}
Since $a/q=6$ is fixed while both grow as $\alpha^2$, increasing $\alpha$ moves along a straight line in the $(a,q)$ plane and crosses higher and higher tongues. The characteristic frequency $\sqrt a=\sqrt3\,\alpha$ grows linearly with $\alpha$, consistent with the growth of the dominant Fourier component, $|\vec K_{mn}|\sim\alpha$. The analogy is only qualitative, since the real problem is two-dimensional and matrix-valued.

\section{Summary of Scales}

\begin{table}[h]
\centering
\caption{Dimensionless scales used in these notes ($v_0=k_\theta=1$).}
\label{tab:scales}
\begin{tabular}{@{}ll@{}}
\toprule
Quantity & Value \\ \midrule
Coupling & $\alpha = w_{AB}/(v_0k_\theta)$, \quad $\theta\approx0.636^\circ/\alpha$ \\
First magic angle & $\alpha_1=0.586$ ($\theta\approx1.09^\circ$) \\
Magic-angle spacing & $\alpha_{m+1}-\alpha_m \to 3/2$; same parity $\to3$ \\
Coherent-state width & $\sigma = 1/\sqrt{3\alpha}$ \\
Localization radius & $|\vec R|\to\pi/3\approx1.05$ ($1.10$ at $\alpha_9$), $0.88$ (radial estimate) \\
Kinetic energy at $\alpha_m$ & $\expval{T}_m=\alpha_m^2\expval{|U|^2}_m\approx\alpha_m^2$ \\
\bottomrule
\end{tabular}
\end{table}

\section*{Acknowledgments}

This work was supported by CONAHCyT project 1564464 and UNAM DGAPA project IN101924. L.A.N.L. was supported by a CONAHCyT PhD scholarship No.\ 1564464. We thank Pierre Anthony Pantale\'on Peralta and Francisco Guinea (IMDEA Nanoscience) for many discussions.

\end{document}